\documentclass[useAMS,usenatbib]{mn2e}

\usepackage{amssymb}
\usepackage{txfonts}
\usepackage{epsfig}

\title[Mass accretion rates of PMS stars in the LMC]{Photometric determination of the mass accretion rates of pre-main sequence stars. III. Results in the Large Magellanic Cloud}
\author[L. Spezzi et al.]{L. Spezzi$^{1}$\thanks{E-mail: lspezzi@rssd.esa.int}\thanks{{\it Facilities:} HST (WFPC2)},  G. De Marchi$^{1}$, N. Panagia$^{2,3,4}$, A. Sicilia-Aguilar$^{5,6}$ and B. Ercolano$^{7,8}$\\
$^{1}$European Space Agency (ESTEC), PO Box 299, 2200 AG Noordwijk, The Netherlands\\
$^{2}$Space Telescope Science Institute, 3700 San Martin Drive, Baltimore, MD 21218\\
$^{3}$INAFÐCT, Osservatorio Astrofisico di Catania, Via S. Sofia 78, I-95123 Catania, Italy\\ 
$^{4}$Supernova Limited, OYV No.~131, Northsound Rd., Virgin Gorda, British Virgin Islands\\ 
$^{5}$Max-Planck-Institut f\"ur Astronomie, K\"onigstuhl 17, 69117 Heidelberg, Germany\\
$^{6}$Departamento de F\'{\i}sica Te\'{o}rica, Universidad Aut\'{o}noma de Madrid, Cantoblanco 28049, Madrid, Spain\\
$^{7}$Ludwig-Maximilians-Universitaet, University Observatory Munich, Scheinerstr.~1, D-81679 M\"unchen, Germany\\
$^{8}$Cluster of Excellence ``Origin and Structure of the Universe'', Boltzmannstr.2, 85748 Garching, Germany}

\begin{document}

\date{Accepted 2011 YYYYY xx. Received 2011 XXXXX yy; in original form 2011 June 27}

\pagerange{\pageref{firstpage}--\pageref{lastpage}} \pubyear{2011}

\maketitle

\label{firstpage}

\begin{abstract}
We present a multi-wavelength study of three star forming regions, spanning the age range 1-14~Myr, located 
between the 30 Doradus complex and supernova SN1987A in the Large Magellanic Cloud (LMC). 
We reliably identify about 1000 pre-main sequence (PMS) star candidates actively undergoing mass accretion 
and estimate their stellar properties and mass accretion rate ($\dot{M}$). 
Our measurements represent the largest $\dot{M}$ dataset of low-metallicity stars presented so far. 
As such, they offer a unique opportunity to study on a statistical basis the mass accretion process in the LMC and, more in general, 
the evolution of the mass accretion process around low-metallicity stars. 
We find that the typical $\dot{M}$ of PMS stars in the LMC is higher than for galactic PMS stars of the same mass, independently of their age. 
Taking into account the caveats of isochronal age and $\dot{M}$ estimates, 
the difference in $\dot{M}$ between the LMC and our Galaxy appears to be about an order of magnitude. 
We review the main mechanisms of disk dispersal and find indications that typically higher $\dot{M}$ 
are to be expected in low-metallicity environments. However, many issues of this scenario need to be clarified by future observations and modeling. 
We also find that, in the mass range 1-2~M$_\odot$, the $\dot{M}$ of PMS stars in the LMC increases with stellar mass as 
$\dot{M}\propto$M$^b _{\star}$ with $b \approx$1, i.e. slower than the second power low found for galactic PMS stars in the same mass regime.
\end{abstract}


\section{Introduction} \label{intro}

The mass accretion rate ($\dot{M}$) of young stars is a key parameter to constrain the models of both star and 
planet formation. $\dot{M}$ is an indicator of the presence of gas, and hence dust, in the inner circumstellar disk and, as such, 
it is affected by the disk structure and evolution as well as by the formation and migration of planets \citep[see, e.g.,][]{Cal00}. 
Thus, it is of particular interest to determine the evolution of $\dot{M}$ as a star approaches its main sequence, whether $\dot{M}$ and stellar mass  are correlated, 
and whether it is affected by the specific conditions of a given star formation event, such as the chemical composition 
and density of the parent molecular cloud, by the proximity of hot massive stars, etc. 
The measure of $\dot{M}$ generally relies on the study of continuum veiling, UV excess emission or 
profile/intensity of emission lines (usually H$\alpha$, Pa$\beta$ or Br$\gamma$), 
which requires medium to high resolution spectroscopy of individual young objects. 
As a result, the data collected so far are limited to nearby (d$\lesssim$1-2~kpc) 
star forming regions. These data indicate that $\dot{M}$ is a decreasing function of 
stellar age \citep{Muz00,Sic05,Sic06,Sic10,Fed10}, in line with the expected 
evolution of viscous discs \citep{Har98}, and is roughly proportional to the second power of the stellar mass \citep{Muz05,Nat06,Sic06}. 
However, the detection limits and the difficulty of measuring 
very small $\dot{M}$ do not allow us to rule out detection/selection thresholds as responsible of this trend \citep[see, e.g.,][]{Cla06}. 
Moreover, the spread of the $\dot{M}$ data exceeds 2~dex at any given age. 
Several facets of the mass accretion process appear to contribute to this spread: i) at any age, disks in different evolutionary stages coexist in the same 
star forming region \citep[see, e.g.,][]{Har07,Alc08,Mer08}; ii) $\dot{M}$ depends on the disk mass and, hence, on the mass of the forming 
star \citep{Whi04,Alc08,Kim09}; iii) the accretion process in young stars, and the accretion-related emission lines, 
are subject to variations on a timescale of a few hours to years or even decades  \citep{Her86,Har91,Ngu09}.
Another limitation of the galactic objects for which $\dot{M}$ measurements are currently available is that they have essentially solar metallicity \citep[e.g.][]{Pad96}, 
because they are located in nearby star forming regions. 

In the past two decades, many authors have demonstrated the potential of deep 
Hubble Space Telescope (HST) optical imaging for star formation studies. 
Recently, \citet{DeM10} (hereafter Paper~I) presented a novel method to identify pre-main sequence (PMS) objects actively undergoing mass accretion and determine their $\dot{M}$. 
This method relies solely on HST imaging, which can be easily obtained for a large number of PMS stars at once and for distant star forming regions, 
well beyond a few kpc and with different metallicity, allowing us to overcome some of the limitations affecting techniques based on spectroscopy. 
This method has been already adapted to the Wide Field Camera (WFC) of the Isaac Newton Telescope (INT) and is being used 
within the frame of INT Photometric H-Alpha Survey (IPHAS), a 1800 deg$^2$ survey of the Northern Galactic Plane 
aiming at reliably select classical T~Tauri stars and  constrain their mass accretion rates \citep{Bar11}. 
We have now started a pilot project aiming at applying this technique to 
the existing and extensive HST photometry of star-forming regions in the 
Milky Way (MW) and the MCs. In this paper we present the results of this 
study in the Large Magellanic Cloud (LMC). 

Because of it relative proximity \citep[51.4$\pm$1.2~kpc;][]{Pan91,Pan91b} and 
low-metal content \citep[Z=0.007;][and references therein]{Mae99}, 
the LMC is a very interesting test-case for star formation studies 
and, specifically, for mass accretion studies, because it offers a unique opportunity to study a resolved 
accreting stellar population in an environment with a chemical composition different from the MW \citep[Z$\approx$0.018; see][]{Est95}. 

Investigating the $\dot{M}$ vs. metallicity dependency is of special interest for several reasons:

\begin{enumerate}

\item The time-scale of the mass accretion process is strongly connected with the disk lifetime, which 
has been poorly studied in low-metallicity environments \citep[see, e.g.,][]{Yas09}. 
Thus, studying the $\dot{M}$ evolution might help clarifying how different disk dispersal mechanisms act at lower-metallicity.

\item Low-metallicity disks can be used as a benchmark for evolved solar-metallicity disks, 
where dust grains may have suffered strong coagulation \citep[see][and discussion at the 
Ringberg Accretion Workshop 2011\footnote{\emph{http://accretion2011.mpe.mpg.de/index.php}}]{Erc10a}. 

\item The evolution of $\dot{M}$ is affected by possible planet formation in the disk and, hence, it 
might provide new important clues on on the so called ``planet-metallicity correlation''. 
The probability of a star hosting a planet appears to increase with stellar metallicity \citep{Gon97,Fis05,San03}. 
However,  a few planets have been found around metal-poor stars \citep[see, e.g.][]{San07,Set03,Set10} 
and the frequency of planets around metal-poor stars compared to solar-metallicity stars is still uncertain. 
Thus, the ``planet-metallicity correlation'' could be a simple selection effect. Indeed, surveys for extra-solar planets are mainly 
based on the radial velocity (RV) method \citep[][and references therein]{May95,Mar98} and RV variations 
are much harder to detect in stellar spectra presenting few metallic lines. 
As a consequence, only a few planets have been found around very metal poor stars and, because of this indication, metal-poor stars are 
systematically excluded from RV surveys.

\end{enumerate}

This paper is organized as follows: Sect.~\ref{data} describes the selection of the LMC fields 
among those available in the HST archive and the data reduction. 
In Sect.~\ref{sel} we apply the method described in Paper~I to select PMS objects and calculate their $\dot{M}$.
The results are presented in Sect.~\ref{results} -\ref{theo}, where in particular we discuss the 
mass dependency of $\dot{M}$, its evolution and metallicity dependency in relation with disk dispersal mechanisms 
such as planet formation and disk photo-evaporation. 
The conclusions of this work are given in Sect.~\ref{conclu}.

\section{Data sample} \label{data}

\begin{figure*}
\includegraphics[angle=0,scale=.7]{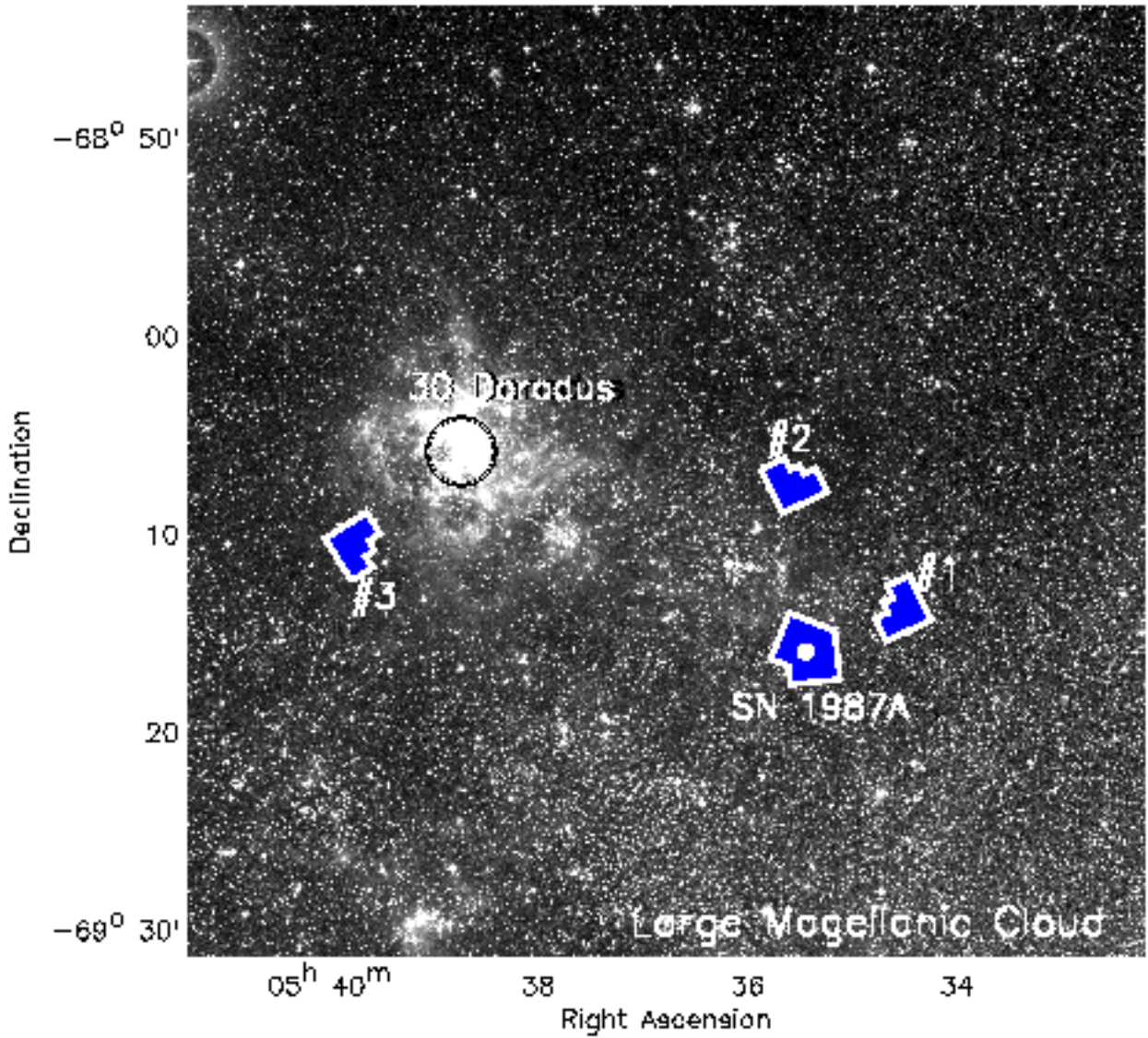}
\caption{R-band image of a 40'$\times$40' area in the LMC form the Sloan Digital Sky Survey. 
The polygons display the four regions observed with HST/WFPC2 analyzed in this paper. 
The position of 30~Doradus star-burst cluster and the supernova SN1987A are also indicated.}
\label{obs}
\end{figure*}

We selected three 162"$\times$162" areas in the LMC located 
between the 30 Doradus star forming complex and supernova SN1987A (Figure~\ref{obs}). 
All fields were selected from the HST Archival Pure Parallel Project\footnote{See http://archive.stsci.edu/prepds/appp for details.}; 
the proposal ID and Principal Investigator (PI) for each dataset are reported in Table~\ref{obs_tab}. 
The observations were performed from February 1999 to November 2003 with the Wide Field Planetary 
Camera 2 (WFPC2) of the HST using the F606W (V), F656N (H$\alpha$) and F814W (I) filters.  
The three fields were selected following two basic criteria: 
i) exposure time long enough to allow the detection of objects down to 0.8-1~M$_{\odot}$ at the 
distance of LMC in the three filters;  ii) different environmental conditions, such as vicinity to OB stars and amount 
of interstellar gas and dust, in order to study the behavior of $\dot{M}$ in star-forming regions with different properties.

The basic data set used for this paper was retrieved from the HST archive 
at the Canadian Astronomy Data Centre (CADC). 
The sources identification and photometry was performed on the stacked images using the SExtractor~2.5 tool by \citet{Ber96}. 
SExtractor exploits the aperture photometry technique, 
which is the fastest and best approach for not-crowded fields as 
those considered here, and the background is locally estimated in an annulus surrounding the star. 
Following the prescriptions of the HST Data Handbook for 
the WFPC2 \citep{Bag02}, we adopted a 0.5~arcsec aperture radius for magnitude extraction and then 
calibrated the instrumental magnitudes to the VEGAMAG photometric system using the 
zeropoints listed in Table~5.1 of the handbook. 
The SExtractor morphological parameter ``extraction flag'' (FLAGS) 
was used to clean the catalogs from spurious detections, such as 
saturated or truncated sources too close to the image boundaries, etc. 
Table~\ref{obs_tab} summarizes, for each of the three fields, the total exposure time, 
the saturation limit and the limiting magnitude achieved at the 10$\sigma$ level in each filter. 
The exposure times are not uniform across our dataset, producing different saturation limits and limiting magnitudes for each filter/region. 
Considering the distance to the LMC ($\sim$50~kpc) and the typical age of our regions (Table~\ref{mean_par}) and using the evolutionary models for PMS stars 
by \citet{Tog11} for the metallicity of the LMC \citep[Z=0.007;][and references therein]{Mae99}, 
we concluded that our observations are complete above 1~M$_\odot$ for the three regions in all the filters. 
Thus, in Sect.~\ref{Macc_mass_age}-\ref{Macc_metal} we will limit our analysis to the mass range 1-2~M$_\odot$.  

In addition to the above fields, we have used WFPC2 photometry 
of a field around supernova SN1987A studied by \citet{Rom98}, 
\citet{Pan00} and \citet{Rom02}. These observations are centered 
at R.A.=5$^h$35$^m$28$^s$ and Dec.=-69$^{deg}$16$^m$12$^s$ 
and cover a field of 9.16 square arcmin. 
The mass accretion rate for the PMS stars in this field were already derived in Paper~I.

\begin{table*}
\begin{center}
\begin{tabular}{ccccccccc}
\hline\hline
Field & HST proposal ID & PI & RAJ2000$_{center}$ & DECJ2000$_{center}$ & Filter & T$_{exp}$ & Mag Sat. & Mag 10$\sigma$  \\
          &                               &      & (hh:mm:ss)                  & (dd:mm:ss)                      &             & (min)         &                   &                                  \\
\hline\hline
\#1       & 8059 &  Stefano Casertano & 05:34:34.03 & -69:14:01.50 & F606W  & 17 & 17.00 & 24.30  \\  
             &            & 				&			&			&     F656N  	& 73 & 15.00 & 21.00  \\
             &             & 				&			&	    		 &	 F814W  & 39 & 17.00 & 23.80  \\
\hline
\#2       & 9634 & James Rhoads & 05:35:34.93 & -69:07:32.36 & F606W  & 49 & 17.50 & 25.20  \\
             &             &			     &			    &          		& F656N  & 68 & 15.50 & 21.30  \\
             &             & 			     &			    &          	        & F814W  & 35 & 16.00 & 24.30  \\  
\hline
\#3        & 8059 & Stefano Casertano & 05:39:40.57 & -69:10:48.43 & F606W  & 8  & 16.00 & 23.80  \\  
             &            &            			&                         & 			& F656N  & 57 & 14.30 & 20.00  \\
             &             &          			&                         &			& F814W  & 15 & 15.50 & 23.50  \\ 
\hline\hline
\end{tabular}
\caption{HST proposal ID, PI name, total exposure time, saturation limit and limiting magnitude at 10$\sigma$ level for the three HST-WFPC2 pointings used in this paper.}
\label{obs_tab}
\normalsize
\end{center}
\end{table*}

\section{Selection of PMS stars}\label{sel}

We select PMS stars in each field and determine their mass accretion 
rates using narrow-band H$\alpha$ (f656n) and broad-band
V (f606w) and I (f814w) dereddend photometry. 
In the following sections we describe the main steps of our selection procedure. 
For a more detailed description of this method we refer the reader to Paper~I. 
 
\subsection{Reddening}\label{reddening}

From a visual inspection of the images, we find evidence of patchy
nebulosity in our fields and we therefore can expect a certain amount of differential reddening.

\begin{figure*}
\includegraphics[angle=0,scale=.6]{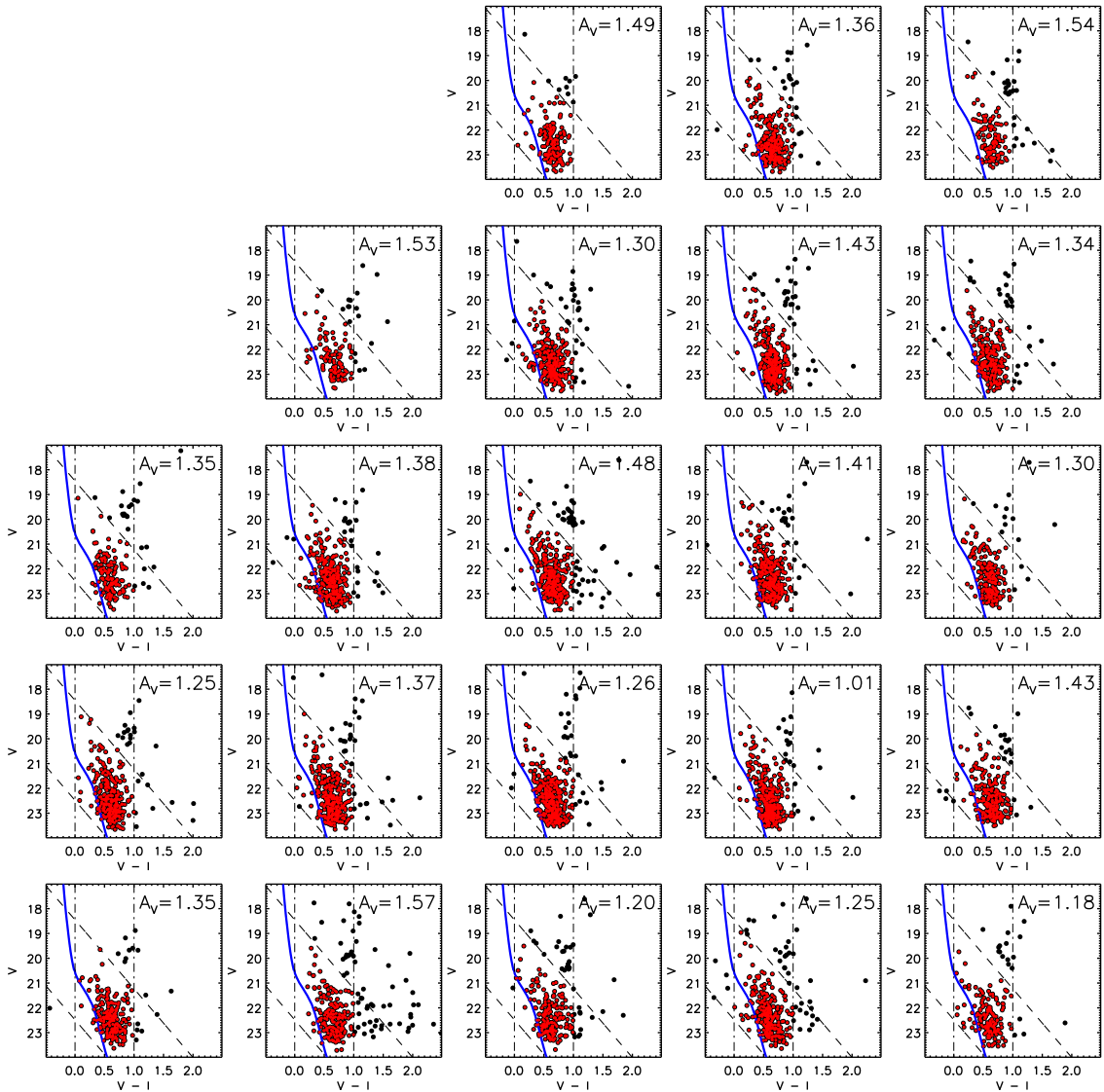} 
\caption{Example $V$ vs. $V-I$ CMDs for one of our WFPC2 fields (field \#1) divided into 
22 32"$\times$32" regions to study the spatial variation of the interstellar extinction over the observed field.
The solid line is the ZAMS by \citet{Mar08}. The dashed lines show the direction of the reddening vector and, together with the dashed dash-dotted lines, 
define the region of the CMD uncontaminated by stars in the red clump. The visual extinction assigned to each region is reported in the top-left of each panel.}
\label{red}
\end{figure*}

To test this hypothesis, we divided each field in 22 regions, each 32"$\times$32" in size, and obtain the V vs. V-I 
colour-magnitude diagram (CMD) for each sub-image. We used the zero age main sequence (ZAMS) by \citet{Mar08} 
for the metallicity of the LMC as the reference unreddened CMD (Figure~\ref{red}). 
The ZAMS was shifted to the distance of the LMC and reddened by the amount corresponding to the intervening absorption along the line
of sight due to the Milky Way, which \citet{Fit84} quantified in A$_V$=0.22, in turn corresponding to E(V ? I) = 0.1.

\begin{figure}
\includegraphics[angle=0,scale=.4]{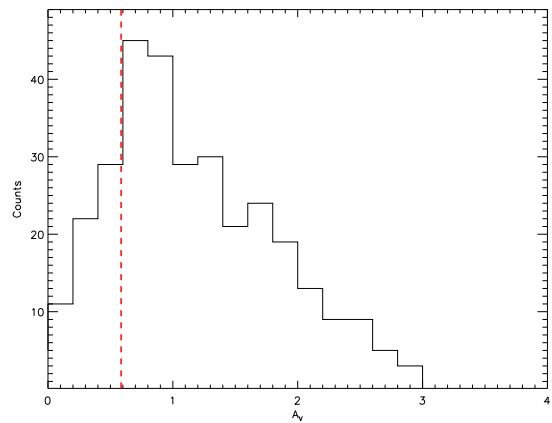} 
\caption{Example of A$_V$ distribution for stars in a sub-image of our field \#1. The dashed line indicates the 17th percentile of the distribution.}
\label{hist_Av}
\end{figure}

Within each sub-image we derived for all stars a single value the reddening relative to the ZAMS in the following way. 
For each star with 18$<$V$<$24 and 0$<$V-I$<$1.3, we calculated the distance from the ZAMS along the reddening vector, 
defined by the extinction law as measured specifically in the LMC by \citet{Scu96}. 
Each one of these distances is the resultant of two components, namely E(V-I) along the abscissa and A$_V$ 
along the ordinates. We then looked at the distribution of A$_V$ values inside each cell and took the 17th percentile 
as a conservative lower limit to the value of A$_V$ for the whole cell (see example in Figure~\ref{hist_Av}). 
If the reddening distribution were Gaussian, this would correspond to 1$\sigma$. 
The purpose of this approach is to minimize the number of stars that after the correction become bluer than 
the ZAMS and to keep them to a fraction compatible with our photometric uncertainties. 
This choice ensures that we are not overestimating the reddening correction for the majority of the
stars in the field. It also avoids an improper overestimation of the ages due to an
excessive reddening correction (see reddening lines in Figure\ref{red}).

The relative A$_V$ values of the sub-images span a range of 0.4~mag in our field \#1, 0.6 mag in our field \#2 and 1.2~mag in our field \#3. 
Thus, differential reddening is more significant in field \#3, which is indeed the closest one 
to the 30~Doradus star-burst cluster and is still associated with conspicuous gaseous/dusty nebulosity (Figure~\ref{spa_dist}).

We correct the magnitudes of each source in each filter assuming the A$_V$ estimated for the relative sub-image 
and the extinction law by \citet{Scu96}. This statistical method, though providing only a rough estimate of the visual extinction 
toward the given source, is suitable for our purposes. Indeed, in Paper~I we have demonstrated that for an average LMC star-forming region 
omitting the extinction correction would result in a 10\% overestimate of the mass accretion rate. 
This error, albeit systematic, is smaller than most other systematic uncertainties and comparable to the measurement errors.

\subsection{Selection of stars with H$\alpha$ excess emission} \label{Ha_sel}

We select PMS stars on the basis of their H$\alpha$ emission line \citep{Whi03}. 
The traditional approach to search for H$\alpha$ emitters
is based on the use of the R-band magnitude as an indicator of the
level of the photospheric continuum near the H$\alpha$ line, 
so that stars with strong H$\alpha$ emission will have 
a large $R-H\alpha$ color. Since we do not dispose of R-band photometry for our fields, 
we use measurements in the neighboring V and I bands, 
following the approach described in Paper~I. 

\begin{figure}
\includegraphics[angle=0,scale=.40]{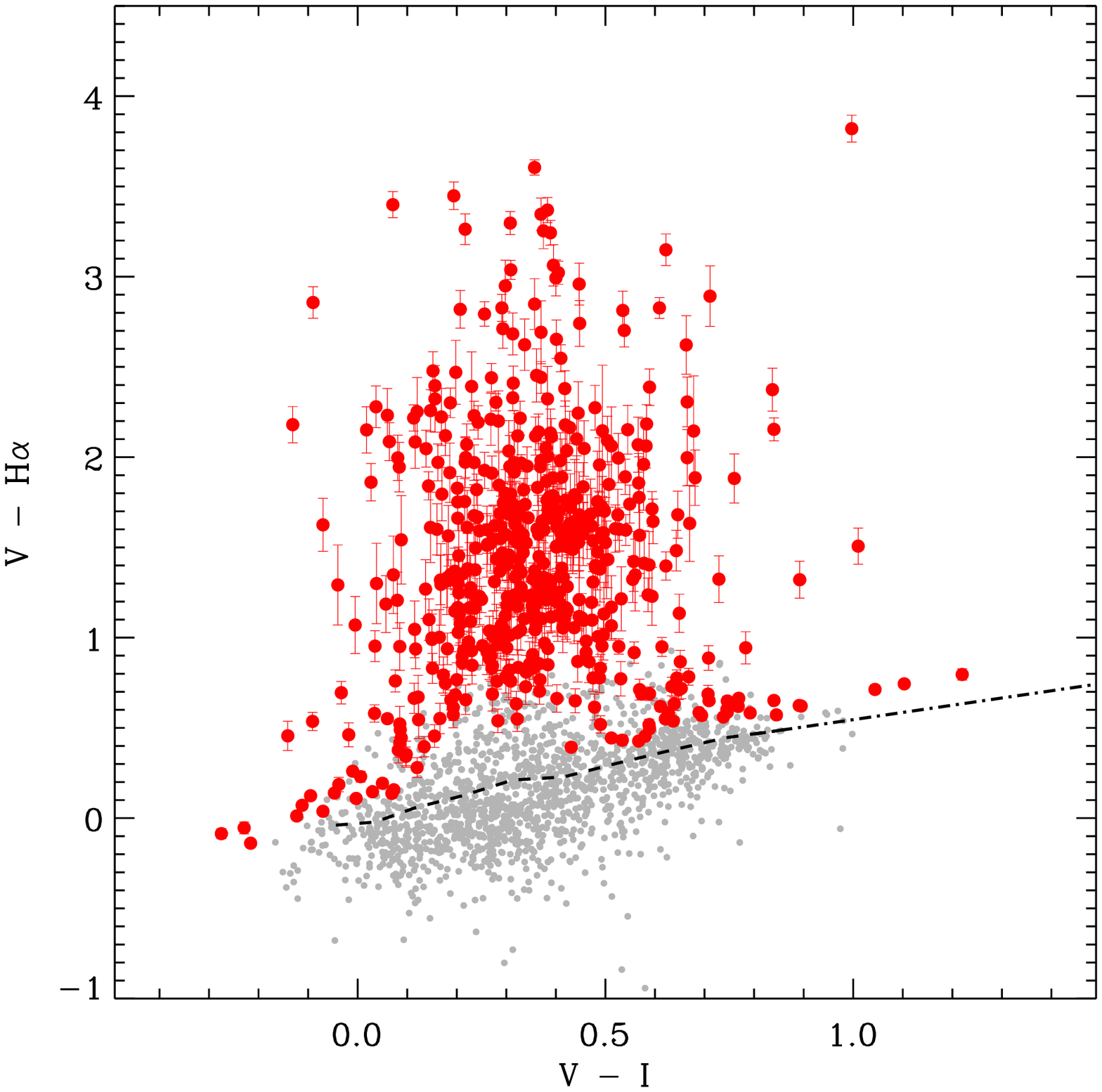}
\includegraphics[angle=0,scale=.40]{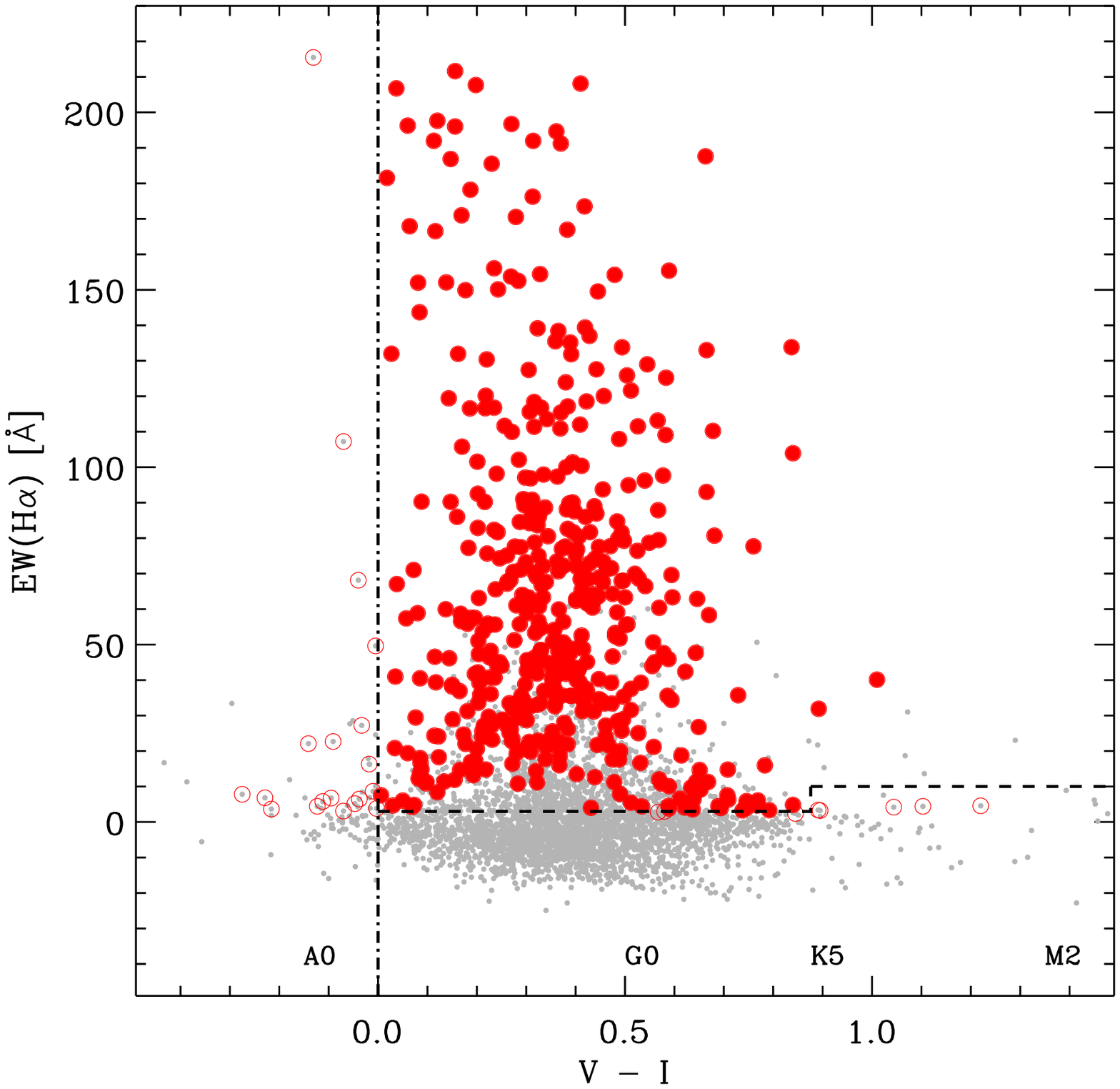}
\caption{{\bf Upper panel:} Example of $V-H\alpha$ vs. $V-I$ diagram of stars in one of our fields (field \#1). 
The line is the reference template with respect to which the excess H$\alpha$ 
emission is sought. Objects with a $V-H\alpha$ excess are 
indicated as big dots with error bars. {\bf Lower panel}: H$\alpha$ equivalent width 
as a function of $V-I$ for stars in field \#1. Objects satisfying the \citet{Whi03} EW$_{H\alpha}$ criterion (dashed line) 
and redder than $V-I$=0 (dot-dashed line) are considered PMS star candidates (big dots).
The remaining objects (open circles) are older stars with chromospheric activity or Ae/Be stars. }
\label{Ha_em}
\end{figure}

Figure~\ref{Ha_em} (top panel) shows, as an example, the $V-H\alpha$ vs. $V-I$ color-color diagram for our field \#1. 
We use the median $V-H\alpha$ dereddened color 
of stars with small photometric errors in the V, I and H$\alpha$ bands ($\delta_3 = \sqrt{(\Delta V^2+\Delta I^2+\Delta H\alpha^2)/3}<0.15$) 
as a function of $V-I$ to define the reference template with respect 
to which the excess H$\alpha$ emission is sought (line in Figure~\ref{Ha_em}). 
Since the contribution of the H$\alpha$ line to the V magnitude is completely negligible, 
the magnitude $\Delta  H\alpha$ corresponding to the excess emission is simply:

\begin{equation} 
\Delta  H\alpha = (V-{\rm H}\alpha)^{\rm obs} - (V-{\rm H}\alpha)^{\rm ref}
\end{equation}

where the superscript "obs" refers to the observations and "ref" to the reference template. 
We select a first sample of stars with excess H$\alpha$ emission by considering
all those with $\delta_3 <$0.15 and V-H$\alpha$ color above the reference template by 5 times
their intrinsic photometric uncertainty ($\Delta  H\alpha > 5 \times \sqrt{\Delta V^2+ \Delta H\alpha^2}$). 
Note that this threshold is somehow arbitrary; it affects the total number of selected PMS star candidates but not 
the typical $\dot{M}$ of the selected PMS population, on which the conclusion of this paper are based (Sect.~\ref{Macc_metal}). 

Then, we derive the equivalent width of the H$\alpha$ emission line (EW$_{H\alpha}$) from the measured $\Delta  H\alpha$ as:
\begin{equation} 
EW_{H\alpha} = RW \times [1-10^{-0.4 \times \Delta  H\alpha}]
\end{equation}
where RW is the rectangular width of the filter, which depends on the characteristics of the filter (see Table~4 in Paper~I). 
We also derive the effective temperature (T$_{eff}$), radius (R$_{\star}$) and luminosity (L$_{\star}$) from the 
dereddened $V-I$ color and V magnitude. We adopt the distance of 51.4~kpc to the LMC 
\citep{Pan91,Pan91b} and use the tabulation of HST/WFPC2 colors as a function of T$_{eff}$ by \citet{Bes98}.

Following the prescription of \citet{Whi03} and adopting the effective temperature vs. spectral type scale by \citet{Ken95}, 
we finally consider as probable PMS stars those objects with (Figure~\ref{Ha_em}, lower panel) EW$_{H\alpha} \geq$3~\AA~for stars earlier than K5 (i.e., T$_{eff} \geq$4350~K), 
EW$_{H\alpha} \geq$10~\AA~for K5-M3 stars (3400$\leq T_{eff} <$4350~K) and 
EW$_{H\alpha} \geq$20~\AA~for M3-M6 stars (3000$\leq T_{eff} <$3400~K). 
Our photometric limits are such that we do not detect LMC members later than M6 in any of the fields (see Sect.~\ref{data}). 
We also reject stars with $V-I \le$0. These two criteria allow us to clean our sample from possible contaminants, 
i.e. older stars with chromospheric activity and Ae/Be stars, respectively. 

Note that dKe and dMe stars with enhanced coronal emission have EW$_{H\alpha}$ well below 20~\AA~ even during flares \citep{Wor76,Bop78}. 
Thus, the criterion by \citet{Whi03} allows us to safely remove dMe stars, while some the dKe might still be included in the PMS sample. 
We could use a more conservative threshold of EW$_{H\alpha}=$20~\AA~ for both K and M type stars, as was done in Paper~II. 
However, while avoiding contamination from magnetically active field stars, this threshold might cause the rejection of 
some of the weaker K0-K7 type accretors \citep[$\dot{M} \lesssim10^{-9} M_{\odot}$/yr;][]{Whi03} and, hence, 
an overestimation of the typical accretion rates of our PMS populations, in particular at ages older than $\sim$10~Myr. 
Thus, for K type stars we choose to adopt  the more relaxed criterion of \citet{Whi03}.

We identified 490 objects in our field \#1, 325 in our field \#2 and 242 in our field \#3 with H$\alpha$ emission typical of accreting PMS stars. 
These numbers are of the same order of magnitude as the number of low-mass stars expected in each region on the basis of 
the number of OB stars (Sect.~\ref{ext_OB_photoevap}) and a typical IMF \citep[$\alpha$=2.35;][]{Sal55}, meaning that 
we are detecting a large fraction of accretors in those regions. 
This point is very important for the conclusions of this paper. As we will see in Sect.~\ref{Macc_metal},  the typical mass accretion rates in the LMC are 
higher than in the MW; because our samples are representative of the entire PMS population in the studied regions, we can be confident that this is not a selection effect.

\subsection{Determination of mass and age} \label{mass_age}

The mass (M$_{\star}$) and age of each star are derived by comparing its L$_{\star}$ and T$_{eff}$ with PMS evolutionary models.  
Specifically, we used the Pisa database of PMS tracks and isochrones \citep{Tog11} for metallicity Z=0.007, as appropriate for the LMC, and 
followed the interpolation procedure developed by \citet{Rom98}, which does not make assumptions 
on the properties of the population, such as the functional form of the IMF. On the basis of the measurement errors, 
this procedure provides the probability distribution for each individual star to have a given value of the mass and age 
(the method is conceptually identical to the one presented recently by \citet{DaR10}).

\begin{figure*}
\begin{center}
\includegraphics[angle=0,scale=.80]{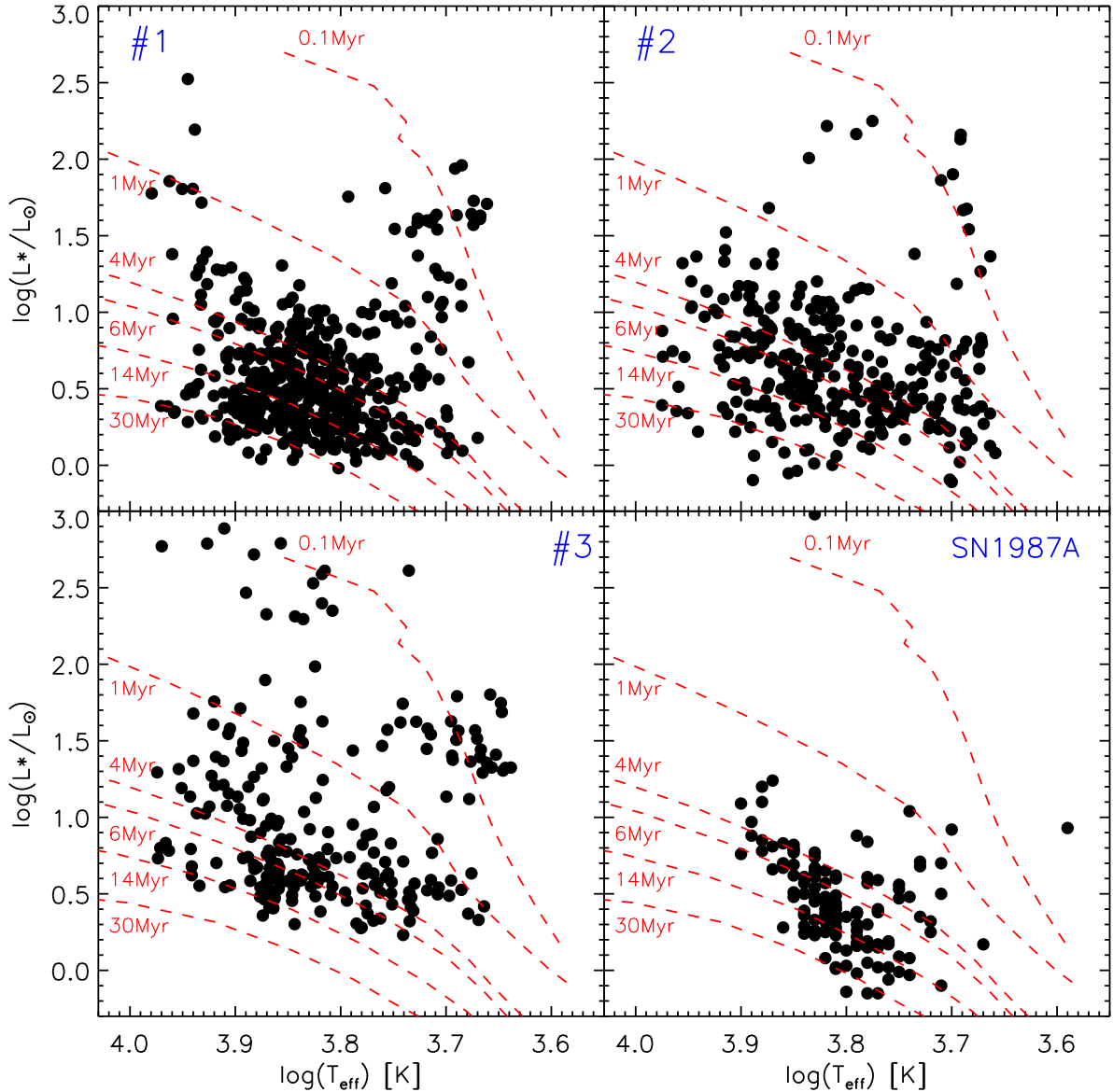}
\caption{HR diagram for the PMS population in the four LMC regions under analysis. 
The dashed lines are the PMS isochrones by \citet{Tog11} for the metallicity of the LMC. 
All objects shown here have H$\alpha$ excess emission and H$\alpha$ equivalent width above the levels discussed in Sect.~\ref{Ha_sel}.}
\label{HR_diag}
\end{center}
\end{figure*}

Many caveats should be considered when dealing with masses and ages of PMS stars derived from evolutionary models. 
We will discuss them in detail in Sec.~\ref{caveats}, because the uncertainty on these parameters, in particular on the isochronal age,  
could have a strong and unavoidable impact on our results.  
Here we stress that, although isochronal ages of individual objects are uncertain, given the presence of unresolved binaries and the stellar variability, 
ages of statistical samples are, in principle, reliable and the global age differences between regions are real \citep[see, e.g.,][]{May07}. 
This statistical approach is the one we use for the analysis illustrated in the next sections.  
Moreover, there are several arguments supporting the reliability of our age estimates:

\begin{enumerate}

\item the median age estimated for our field \#1 (13~Myr, Table~\ref{mean_par}), 
the closest one to supernova SN1987A, is roughly consistent with the age independently estimated for the progenitor of SN1987A 
and nearby stars \citep[$\sim$13~Myr;][]{Scu96};

\item Figure~\ref{HR_diag}  shows the HR diagram for the PMS populations of the four LMC regions under analysis.
There is a clear difference between the distribution of PMS stars belonging to the youngest (\#3, 6-8~Myr) and the oldest (SN1987A, 13~Myr) of our regions, 
with region \#3 being clearly a few Myr younger than the SN1987A region. 
The age of regions \#1 and \#2 is intermediate between region \#3 and the SN1987A region and their PMS population concentrate 
between 4 and 14 Myr, where the isochrones become very close to each other and it is hard to 
derive an age difference based only on the visual inspection of the HR diagram. 
However, as reported in Table~\ref{mean_par}, the median isochronal ages of region \#1 and \#2 differ by a few Myr 
and with respect to region \#3 and the SN1987A region as well, and the difference is statistically significant. 

\item \citet{Elm00} showed that the age dispersion of a given stellar population depends on the size of the region. 
Taking into account the typical size of our region (162"$\times$162", i.e.  about 100~pc at the LMC distance), we would expect an age dispersion of a 
few to 10 Myr inside each region, which is consistent with our findings (Table~\ref{mean_par}). 

\end{enumerate} 

\subsection{Determination of the mass accretion rate}\label{par_Macc}

The mass accretion rate is derived from the free-fall equation as:

\small
\begin{equation}
\label{equ_Macc}
\log \dot{M} (M_{\odot}/yr)= -7.39 + \log \frac{L_{acc}}{L_{\odot}} + \log \frac{R_{\star}}{R_{\odot}} - \log \frac{M_{\star}}{M_{\odot}}
\end{equation} 
\normalsize
where L$_{acc}$ is the energy released by the accretion process, directly
proportional to the H$\alpha$ luminosity \citep[][and Paper~I]{Dah08}:

\begin{equation}
\label{equ_Lacc}
\log L_{acc}=(1.72 \pm 0.47) + \log L(H\alpha)
\end{equation}
Once the color excess $\Delta H\alpha$ is determined, the H$\alpha$ 
emission line luminosity, L(H$\alpha$), can be immediately
obtained from the photometric zero point and absolute
sensitivity of the instrumental set-up and from the distance
to the sources (see Sect.~3.1 in Paper~I).

The problems involved in measuring $\dot{M}$ have been recently discussed by several authors \citep[e.g.,][]{Her08,Bar11,Rig11} and 
measurements using different diagnostics might give different results by about an order of magnitude. 
The uncertainties of  the H$\alpha$ diagnostics and the specific method adopted in this work have been fully addressed in Paper~I. 
The typical error on $\log{\dot{M}}$ is of the order of 0.7-0.8 dex (see Table~\ref{tab_par}) and the main sources of this uncertainty are:

\begin{enumerate} 

\item  The uncertainty on the L$_{acc}$ vs. L(H$\alpha$) calibration relation (Equation~\ref{equ_Lacc}), which is based on near-infrared spectra of a dozen 
accreting members of the IC~348 cluster \citep[see Figure~17 by][]{Dah08}. Due to the poor statistics, the resulting uncertainty on L$_{acc}$ is as high as a factor of 3. Moreover, 
the relation is mainly calibrated using stars with mass below 1~M$_\odot$ and a handful of stars around 2~M$_\odot$, while a gap in accreting stars is observed 
extending between 0.8 and 1.8~M$_\odot$. Thus, this relation is more uncertain when applied to the mass regime considered in this paper (1-2~M$_\odot$);

\item Systematic uncertainties due to i) discrepancies in the isochrones and evolutionary tracks, ii) reddening, 
iii) H$\alpha$ emission not due to the accretion process, and iv) the contribution of the nebular continuum to the colors of the stars. 
In Paper~I we discussed in detail these four contributions and concluded that:
i) systematic uncertainties of the order of 20\% are to be expected for the mass due to differences between evolutionary models; 
ii) for an average LMC star-forming region, omitting the extinction correction would result in a 10\% overestimate of $\dot{M}$. 
This error is smaller in our case, because our photometric catalogs are dereddened taking into account the effect of patchy absorption (Sect.~\ref{reddening}); 
iii) possible sources of H$\alpha$ emission other than accretion are chromospheric activity, whose contribution is 2 orders of magnitude less than what we measure, 
H$\alpha$ emission along the line of sight arising in knots of ionized H (e.g., very compact HII), which we removed  by a visual inspection 
of all objects with excess H$\alpha$ emission, and gaseous regions around the object ionized by an external source (e.g., hot ionizing stars), 
which, according to the calculation presented in Paper~I (Table~1), is negligible in our case because there are no OB stars in the close vicinity ($<$5") 
of our PMS star candidates (see Sect.~\ref{ext_OB_photoevap}). 
We can, therefore, conclude that in our fields the contribution of H$\alpha$ emission generated by sources other than the accretion is negligible; 
iv) if present, nebular continuum will add to the intrinsic continuum of the star, thereby affecting the observed broad-bands colors of the source. 
As shown in Paper~I (Table~2), for the typical spectral range of our candidates (G-K), the effects of the nebular continuum on the $V-I$ 
color of PMS stars remains insignificant even for the PMS objects with the highest  EW$_{H\alpha}$;

\item Other possible systematic errors on the mass and age estimates due to inaccuracies in the models' input physics, which we will discussed in Sect.~\ref{caveats};

\item As for the statistical uncertainty on other quantities in Equation~\ref{equ_Lacc} and \ref{equ_Macc}, they are as follows. 
With our selection criteria, the typical uncertainty on L(H$\alpha$) is 15\% and is dominated by random errors. 
The uncertainty on R$_\star$ is typically 7\%, including a 5\% systematic uncertainty on the distance modulus. As for the mass M$_\star$, 
since it is determined by comparing the location of the star in the HR diagram with evolutionary tracks, both systematic and statistical uncertainties are important. 
The uncertainty on the temperature is mostly statistical and is about 3\%, while that on the luminosity is 7\%, 
comprising both random errors (1\% uncertainty on the bolometric correction and 3\% on the photometry) and systematic effects (5\% uncertainty on the distance modulus). 
When we interpolate through the PMS evolutionary tracks to estimate the mass, the uncertainties on T$_{eff}$ and L$_\star$ imply an error of ~7\% on M$_\star$. 
In summary, the combined statistical uncertainty on $\dot{M}$ is 17\%.

\end{enumerate}

\section{The new sample of accreting PMS star candidates in the LMC}\label{results}

We have determined the mass accretion rates for 490 PMS star candidates in our field \#1, 325 in our field \#2 and 242 in our field \#3.
Adding the 106 PMS stars  in the field around supernova SN1987A characterized in Paper~I, 
we end up with a sample of about 1000 PMS star candidates in the LMC with mass
between 1 and 5~M$_{\odot}$ and age between $\sim$1 and 30~Myr. Our sample 
represents the largest $\dot{M}$ dataset for low-metallicity stars presented so far. 
The majority of our PMS stars ($\sim$90\%) have masses below 2~M$_\odot$, i.e. in the T~Tauri star range. 
The remaining $\sim$10\% are Herbig Ae/Be stars; at the typical age of our regions (Table~\ref{mean_par}), many of them 
could already approach or be on the main sequence. Moreover, they cannot be considered simply as massive counterparts of T~Tauri stars 
because they are characterized by different properties \citep[strong winds, small magnetic fields, etc.;][]{Her60,Wat98}. 
Thus, the H$\alpha$ vs. $\dot{M}$ calibration used for T~Tauri stars (Equation~\ref{equ_Lacc}) might be inappropriate for Herbig Ae/Be stars and, hence, 
the $\dot{M}$ derived for these objects might be inaccurate. This does not affect the results presented in this paper 
because, as explained in Sect.~\ref{data}, we limit our analysis to the range 1-2~M$_\odot$.

HST/WFPC2 photometry, stellar parameters and mass accretion rates for this sample are available, 
only in electronic form, in Table~\ref{tab_phot} and \ref{tab_par}, respectively.  
In Table~\ref{mean_par} we summarize the number of PMS stars selected in each field and their median mass, age and mass accretion rate. 
Using this large sample of PMS stars with $\dot{M}$ measured in a homogeneous way, 
we now investigate the dependency of $\dot{M}$ on stellar mass, age and metallicity, in relation with the proposed mechanisms of disk dispersal.

\begin{table*}
\begin{center}
\caption{Astrometry and observed photometry of the PMS star candidates in the LMC.}
\begin{tabular}{ccccccc}

\hline\hline
Field &     ID   & RAJ2000     &  DECJ2000 & F606W & F656N & F814W \\
          &            & (hh:mm:ss)  & (dd:mm:ss)   &	        &              &              \\ 
\hline\hline
\#1  &    1  &   5:34:47.62  &  -69:14:37.89  &  22.435$\pm$0.028  &  20.463$\pm$0.109  &  21.957$\pm$0.028  \\  
\#1  &    2  &   5:34:44.36  &  -69:15:13.14  &  21.028$\pm$0.012  &  19.631$\pm$0.051  &  20.220$\pm$0.010  \\  
\#1  &    3  &   5:34:43.90  &  -69:15:17.41  &  21.549$\pm$0.015  &  20.408$\pm$0.092  &  21.061$\pm$0.015  \\  
\#1  &    4  &   5:34:45.31  &  -69:15: 1.14  &  19.870$\pm$0.008  &  18.990$\pm$0.027  &  19.142$\pm$0.007  \\  
\#1  &    5  &   5:34:47.21  &  -69:14:41.45  &  19.789$\pm$0.008  &  18.839$\pm$0.024  &  18.984$\pm$0.006  \\  
\hline
\#2  &    1  &   5:35:41.65  &  -69: 9:10.82  &  20.484$\pm$0.002  &  18.439$\pm$0.017  &  19.527$\pm$0.002  \\  
\#2  &    2  &   5:35:41.40  &  -69: 9: 8.94  &  23.139$\pm$0.035  &  20.382$\pm$0.092  &  22.338$\pm$0.038  \\  
\#2  &    3  &   5:35:41.55  &  -69: 9: 8.70  &  22.368$\pm$0.016  &  20.311$\pm$0.103  &  21.878$\pm$0.018  \\  
\#2  &    4  &   5:35:41.69  &  -69: 9: 8.25  &  22.095$\pm$0.009  &  20.610$\pm$0.111  &  21.569$\pm$0.012  \\  
\#2  &    5  &   5:35:41.93  &  -69: 9: 8.08  &  21.438$\pm$0.005  &  19.892$\pm$0.060  &  20.738$\pm$0.006  \\  
\hline
\#3-Pop.1  &    1  &   5:39:32.88  &  -69: 9:34.32  &  20.324$\pm$0.011  &  16.878$\pm$0.013  &  19.452$\pm$0.009  \\  
\#3-Pop.1  &    2  &   5:39:36.03  &  -69: 9: 8.81  &  20.434$\pm$0.011  &  19.061$\pm$0.060  &  19.432$\pm$0.010  \\  
\#3-Pop.1  &    3  &   5:39:32.47  &  -69: 9:40.29  &  21.993$\pm$0.036  &  18.877$\pm$0.067  &  21.842$\pm$0.034  \\  
\#3-Pop.1  &    4  &   5:39:32.94  &  -69: 9:36.18  &  20.604$\pm$0.012  &  19.311$\pm$0.082  &  20.331$\pm$0.014  \\  
\#3-Pop.1  &    5  &   5:39:35.28  &  -69: 9:16.37  &  21.993$\pm$0.024  &  20.400$\pm$0.134  &  21.354$\pm$0.021  \\  
\hline\hline
\end{tabular}
\label{tab_phot}
\end{center}
\end{table*}

\begin{table*}
\begin{center}
\caption{Stellar parameters, H$\alpha$ equivalent width and accretion rate of the PMS star candidates in the LMC. }
\begin{tabular}{cccccccc}
\hline\hline
Field &  ID &  $log{T_{eff}}$ &   $log{L_{\star}/L_{\odot}}$ &   Mass	              &   Age	   &    EW$_{H\alpha}$ &   $\log{M_{acc}}$  \\
      &     &     (K)         & 		             &  (M$_{\odot}$)         &  (Myr)     &        (\AA)      & (M$_{\odot}$/yr)   \\ 
\hline\hline
\#1  &    1  &  3.86$\pm$0.02  &  0.55$\pm$0.09  &   1.3  &  17.7  &    67$\pm$ 9  &  -7.4$\pm$0.8  \\ 
\#1  &    2  &  3.72$\pm$0.01  &  1.10$\pm$0.04  &   2.3  &   0.6  &    18$\pm$ 2  &  -7.1$\pm$0.8  \\ 
\#1  &    3  &  3.86$\pm$0.01  &  0.90$\pm$0.05  &   1.5  &   9.2  &    16$\pm$ 3  &  -7.6$\pm$0.8  \\ 
\#1  &    4  &  3.75$\pm$0.00  &  1.54$\pm$0.03  &   3.2  &   0.7  &     4$\pm$ 0  &  -7.2$\pm$0.8  \\ 
\#1  &    5  &  3.73$\pm$0.00  &  1.61$\pm$0.03  &   3.5  &   0.3  &     3$\pm$ 0  &  -7.2$\pm$0.8  \\ 
\hline
\#2  &    1  &  3.66$\pm$0.00  &  1.37$\pm$0.02  &   1.2  &   0.0  &    58$\pm$ 1  &  -5.8$\pm$0.7  \\ 
\#2  &    2  &  3.71$\pm$0.02  &  0.24$\pm$0.12  &   1.3  &   5.9  &   158$\pm$16  &  -7.2$\pm$0.8  \\ 
\#2  &    3  &  3.84$\pm$0.01  &  0.53$\pm$0.06  &   1.3  &  16.4  &    73$\pm$ 9  &  -7.4$\pm$0.8  \\ 
\#2  &    4  &  3.83$\pm$0.01  &  0.63$\pm$0.04  &   1.3  &  12.0  &    31$\pm$ 6  &  -7.6$\pm$0.8  \\ 
\#2  &    5  &  3.75$\pm$0.00  &  0.89$\pm$0.03  &   2.0  &   2.7  &    32$\pm$ 3  &  -7.2$\pm$0.7  \\ 
\hline
\#3-Pop.1  &    1  &  3.68$\pm$0.00  &  1.36$\pm$0.04  &   1.4  &   0.1  &   292$\pm$ 5  &  -5.3$\pm$0.7  \\ 
\#3-Pop.1  &    2  &  3.67$\pm$0.00  &  1.44$\pm$0.04  &   1.4  &   0.0  &    17$\pm$ 2  &  -6.4$\pm$0.8  \\ 
\#3-Pop.1  &    3  &  3.97$\pm$0.04  &  0.73$\pm$0.19  &   1.6  &   8.5  &   245$\pm$19  &  -7.0$\pm$0.8  \\ 
\#3-Pop.1  &    4  &  3.91$\pm$0.01  &  1.21$\pm$0.04  &   1.8  &   6.2  &    21$\pm$ 3  &  -7.2$\pm$0.8  \\ 
\#3-Pop.1  &    5  &  3.80$\pm$0.01  &  0.73$\pm$0.08  &   1.4  &   8.5  &    29$\pm$ 7  &  -7.4$\pm$0.8  \\ 
\hline\hline
\end{tabular}
\label{tab_par}
\end{center}
\end{table*}

\begin{table*}
\caption{Number of PMS stars selected in each field and their median mass, age and mass accretion rate. 
For each parameter the range spanned by each PMS  population and the standard deviation of the distributions ($\sigma$) are also given.}
\begin{center}
\begin{tabular}{cccccc}
\hline\hline
Field & N. of PMS stars & Median Mass    & Median Age   & Median $\log{\dot{M}}^\dag$ &  Corrected $\log{\dot{M}}^\ddag$ \\
         &                              &  (M$_{\odot}$)   & (Myr)               &      (M$_{\odot}$/yr)                   &  (M$_{\odot}$/yr)                              \\ 
\hline\hline
\#1                        & 490            & 1.3	                       & 13                           & -7.5			  &	-7.6	\\
                              &                   & range [1: 5]       & range [1: 30]             &   range [-8.1: -6.3]   &         	\\
                             &                    & $\sigma$=0.5        & $\sigma$=6           &  $\sigma$=0.3        &          	\\
\hline
\#2                       & 325           & 1.4	                     & 9	                             & -7.3			  &	-7.4	\\
                            &                   & range [1: 5]            & range [1: 30]         & range [-7.7: -5.3] 	     & 	        \\
                            &                    & $\sigma$=0.5        & $\sigma$=6           &  $\sigma$=0.5           &       	\\
\hline
\#3 - Pop.~1     & 96              &  1.4	                   &  8                         & -7.1				  &	-7.2	\\  
                           &                    &  range [1: 4]        & range [1: 25]      & range [-7.5: -5.1]       	  &	         \\
                           &                    & $\sigma$=0.4      & $\sigma$=6        &    $\sigma$=0.5          	  &		\\
\hline
\#3 - Pop.~2     & 146            &  1.5	                   &  6                        & -7.1				  &	-7.2	\\  
                           &                    &  range [1: 5]        & range [1:30]        & range [-7.9: -5.3]      	  &		 \\
                           &                    & $\sigma$=1        & $\sigma$=6        &  $\sigma$=0.5                 &		 \\
\hline
SN1987A$^a$ & 104              & 1.3                         &  14	                   & -7.59		                & --	\\  
             	        &                       & range [1: 2.2]       &  range [1: 30]        & range [-8.0: -6.4]          &   	\\
                           	&                      & $\sigma$=0.2        & $\sigma$=6        &  $\sigma$=0.3             &           \\
\hline\hline
\end{tabular}
\label{mean_par}
\normalsize
\end{center}
$^a$\footnotesize{From Paper~I.}\\
$^\dag$\footnotesize{Median mass accretion rate in the mass range 1-2~M$_\odot$.}\\
$^\ddag$\footnotesize{Median mass accretion rate in the mass range 1-2~M$_\odot$ corrected for completeness effects (Sect.~\ref{complet}).}\\
\end{table*}

\subsection{On the completeness of our accreting PMS star sample}\label{complet}

In Sect.~\ref{data} we estimated that our observations for field \#1, \#2 and \#3 are photometrically complete 
in all the filters in the mass range 1-2~M$_\odot$ for 1 to 14~Myr old objects at the distance of the LMC and, hence, 
in this section we limit our analysis to this mass range. 
However, the mass accretion rate of PMS stars depends on the stellar mass \citep{Muz05,Nat06,Sic06} and decreases 
with time \citep{Muz00,Sic05,Sic06,Sic10,Fed10}. Thus, our limit for the measurement of $\dot{M}$ varies within the considered mass and age range. 

\begin{figure*}
\begin{center}
\includegraphics[angle=0,scale=.60]{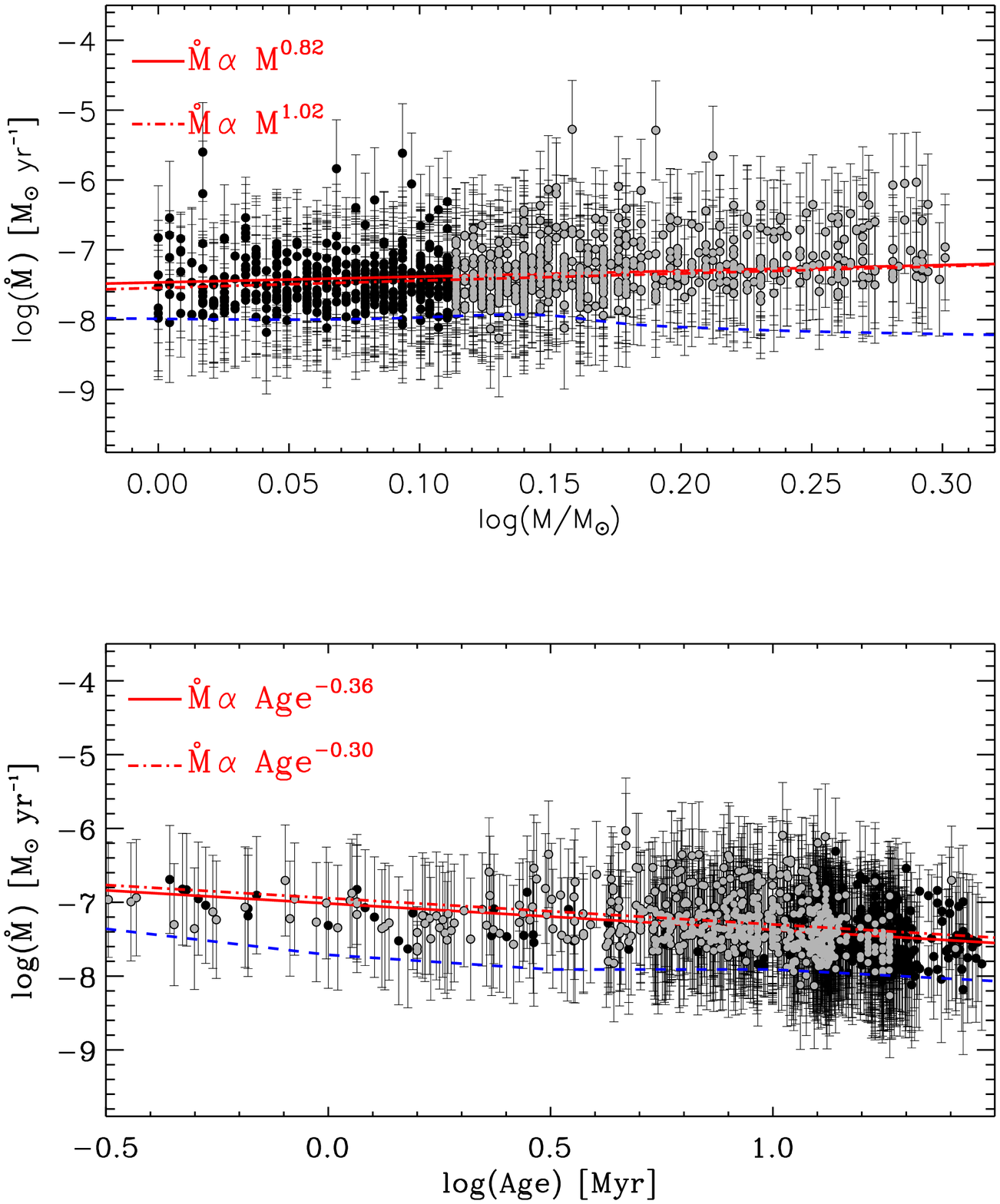}
\caption{Mass accretion rate as a function of the stellar mass (upper panel) and age (lower panel) for PMS star candidates in the LMC 
with masses in the range 1-2~M$_\odot$. In both panels, the solid line displays the result of the double linear regression fit to 
the points obtained by setting M$_{\star}$ and age as independent variables and $\dot{M}$ as dependent variable, 
while the dashed line  represents our photometric limit for the 
measurement of $\dot{M}$ (Sect.~\ref{Macc_mass_age} and \ref{Macc_metal}). 
The gray dots are the data points in the M$_\star \gtrsim 1.3 M_\odot$ mass range, where our PMS star candidate sample is photometrically complete. 
The dot-dashed line displays the result of the double linear regression fit of this sub-set of data. }
\label{fig_Macc_mass_age}
\end{center}
\end{figure*}

Figure~\ref{fig_Macc_mass_age} shows the $\dot{M}$ of PMS star candidates in our sample as a function of mass (upper panel) and age (lower panel). 
In each panel, the dashed line displays our photometric limit for the detection of excess H$_\alpha$ emission and, hence, for the measurement of $\dot{M}$. 
In order to translate the detection limit in H$_\alpha$ luminosity into the detection limit in $\dot{M}$ as a function of stellar mass, 
we considered the lowest H$_\alpha$ luminosity measured for our PMS stars and 
used Equ.~\ref{equ_Macc} and \ref{equ_Lacc}. We assumed the median age of all PMS star candidates in our sample (10~Myrs) and calculated the corresponding 
stellar radii for different stellar masses using the 10~Myr isochrone for the metallicity of the LMC. 
Analogously, we translated the detection limit in H$_\alpha$ luminosity into the detection limit in $\dot{M}$ as a function of stellar age by 
using Equ.~\ref{equ_Macc} and \ref{equ_Lacc} and assuming the lowest H$_\alpha$ luminosity measured for 
our PMS star candidates and their median mass (1.4~M$_\odot$); we then calculated the corresponding stellar 
radii from the 1.4~M$_\odot$ evolutionary track for the metallicity of the LMC. 

Figure~\ref{fig_Macc_mass_age} and the limiting magnitudes in Table~\ref{obs_tab}
show that our photometric limits allow us to detect only the brightest and, hence, youngest accreting late K-type stars at the distance of the LMC, 
while they are sufficient to detect early K or G-type accreting stars (M$_\star \gtrsim 1.3 M_\odot$) of any age up to the main sequence. 
In particular, the lower panel of Figure~\ref{fig_Macc_mass_age} shows that our PMS star candidates define a bulge of points 
at ages around 10~Myrs and this bulge is very close to the photometric limit; this implies that we might have missed part of 
the PMS objects with very low accretion rates at ages around 10~Myr. 
At younger ages, the mass accretion process of PMS stars is enhanced and the H$_\alpha$ luminosity increases; 
if more accreting PMS stars younger than 10~Myr were present, we would have them. 
In summary, our accreting PMS star candidate sample in the LMC is complete in the spanned age range for M$_\star \gtrsim 1.3 M_\odot$, 
while we might have missed part of the oldest lower-mass accreting stars. 

We now quantify this incompleteness effect by using the dataset presented in Paper~I as a reference.
Indeed, one of the main objectives of this work is to compare the accreting properties of PMS stars in our LMC fields with the 
pioneer study presented in Paper~I, where we found that the $\dot{M}$ of PMS stars in the region around supernova SN1987A 
are systematically higher than their Galactic counterparts (Sect.~\ref{Macc_metal}). 
To make sure that this comparison is performed in a homogeneous way, 
we derived a correction to be applied to the typical $\dot{M}$ measured in our LMC fields to match the completeness of the Paper~I dataset. 
We proceeded as follows.

\begin{figure}
\begin{center}
\includegraphics[angle=0,scale=.4]{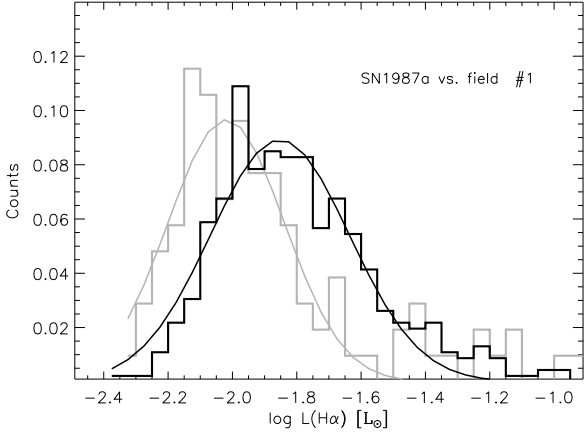}
\caption{Comparison between the H$\alpha$ luminosity distribution of the PMS population in the 
region around SN1987A (gray histogram, from Paper~I) and the distribution for our fields \#1 (black histogram). 
The curves show the Gaussian fit to the data.}
\label{LHa_hist}
\end{center}
\end{figure}

Since the photometric catalog for the SN1987A region (Paper~I) is much deeper than our catalogs, 
it can be considered as a reference catalog with 100\% completeness for the measurement of $\dot{M}$ in the considered mass range (1-2~M$_\odot$). 
We considered the H$\alpha$ luminosity ($L(H\alpha)$) distribution of PMS stars in the field around SN1987A and derived 
its peak by performing a Gaussian fit (Figure~\ref{LHa_hist}). This peak  ($L(H\alpha)_{Ref}$) represents the limit of this dataset for the detection 
of H$\alpha$ excess emission and, hence, for the measure of $\dot{M}$.

In Figure~\ref{LHa_hist} we compare the $L(H\alpha)$ distribution of the SN1987A PMS population 
with the analogous distributions for our region \#1. This region is very close to SN1987A, has the same age (Table~\ref{mean_par}) 
and is part of the same star-forming complex. Thus, he difference between the peak of this distribution and $L(H\alpha)_{Ref}$
represent the $L(H\alpha)$ correction to be applied to field \#1 to match the completeness of the SN1987A dataset. 
The  $L(H\alpha)$ correction has been translated into $\dot{M}$ correction by using Equ.~\ref{equ_Macc} and \ref{equ_Lacc}, 
assuming the median mass and age of the PMS population in region \# as reported in Table~\ref{mean_par} and using the relative 
evolutionary model for the metallicity of the LMC to calculate the corresponding stellar radius.

The H$\alpha$ observations of fields \#2 and \#3 have roughly the same photometric depth as field \#1 (Table~\ref{obs_tab}) and, 
hence, we assume the same $\dot{M}$ correction. 
The median $\dot{M}$ of the PMS star candidates selected in each field corrected for completeness are reported in Table~\ref{mean_par} .

\section{On the mass dependency of the mass accretion rate in the LMC}\label{Macc_mass_age}

The mass accretion rate of galactic PMS stars appears to be roughly proportional to the second 
power of the stellar mass over a wide range \citep[0.3-3~M$_{\odot}$;][]{Nat04,Moh05,Nat06,Muz05,Sic06}, with a large spread 
of two orders of magnitude at least for any interval of M$_\star$. 
Recent observations also show that this power law might even be steeper ($\dot{M} \propto M^{3.1} _{\star}$) 
in the very low-mass and sub-solar mass domain (M$_{\star} \lesssim 0.4 M_{\odot}$), 
while flattening ($\dot{M} \propto M^2 _{\star}$ or even flatter) when only the M$_{\star} \gtrsim 0.8 M_{\odot}$ domain is considered \citep{Fan09,Sic10}. 
This seems to support the idea that the accretion and disk evolution process might present important differences for low and high-mass stars \citep{Har06}.

The large spread observed in the $\dot{M}$ vs.  $M _{\star}$ dependency partly arises from the uncertainty in the determination of $\dot{M}$, 
which can be as high as one order of magnitude. However, this relation still contains an intrinsic scatter 
due to the age dependency of $\dot{M}$, individual disk properties and the variability 
of the mass accretion process, as discussed in Sect.~\ref{intro}.
Indeed, mass accretion rates for younger stars tend to be higher, 
as expected from the viscous disk evolutionary models \citep{Har98}.  
The $\dot{M}$ versus age dependency appears to be very steep, with $\dot{M}$ decreasing 
by more than three order of magnitude within the first 
10~Myr of the stellar life \citep[see Figure~2 by][]{Sic10}. 
This effect is likely to increase the spread in $\dot{M}$ for any interval of M$_\star$, 
because the PMS populations in young clusters and star forming regions 
in the MW generally present an age spread of a few Myr \citep{Jef11}, and objects progressively younger are expected to form 
parallel trends in the $\dot{M}$ vs.  M$_{\star}$ relation with increasing offset in $\dot{M}$ \citep[see, e.g.,][]{Sic06,Sic10}.

We now use our sample of PMS stars to investigate the $\dot{M}$ versus M$_{\star}$ correlation in the LMC. 
Although our sample spans a wide range of stellar masses (see Table~\ref{mean_par}), 90\% of the PMS stars have masses between 1 and 2~M$_\odot$ and, thus, 
we limit our analysis to this range, in order to obtain statistically robust results.  
We find that best fit to the $\dot{M}$ versus M$_{\star}$ dependency in the LMC is a linear correlation  (Figure~\ref{fig_Macc_mass_age}, upper panel). 
In order to investigate quantitatively this dependency, disentangling the dependency on age, 
we performed a double linear regression fit, setting M$_{\star}$ and age as independent 
variables and $\dot{M}$ as dependent variable. In a logarithmic scale, the functional form of this regression is as follows:

\begin{equation}
\label{regr}
\log \frac{\dot{M}}{M_{\odot} yr^{-1}} =  a \cdot \log Age (Myr) + b \cdot \log \frac{M_{\star}}{M_{\odot}} + c
\end{equation}

where $c$ is the offset and $b$ and $a$ the coefficients enclosing the mass and age dependency, respectively. 
In this section, we focus our attention of the mass dependency of $\dot{M}$ enclosed in the $b$  parameter. 
The age dependency and the relative $a$ parameter will be discussed in the Sect.~\ref{Macc_metal}.

Considering all the PMS stars in the range 1-2~M$_\odot$, we find that 
the $\dot{M}$ of PMS stars in the LMC increases almost linearly with stellar mass ($b =0.82 \pm 0.16$, i.e. $\dot{M}\propto$M$_{\star}$). 
The solid line in Figure~\ref{fig_Macc_mass_age} (upper panel) displays the results of our linear regression fit; 
the dashed line displays our photometric limit for the detection of excess H$_\alpha$ emission and, hence, for the measurement of $\dot{M}$ (Sect.~\ref{complet}). 
In order to avoid the mass regime where our survey is not complete, we recomputed the coefficients in Equ.~\ref{regr} limiting 
the regression fit to the M$_\star \gtrsim 1.3 M_\odot$ range (grey points and dot-dashed line in Figure~\ref{fig_Macc_mass_age}). 
In this case, we obtain $b=1.02 \pm 0.31$, confirming that the $\dot{M}$  vs. M$_\star$ dependency is roughly linear.

We conclude that in the approximate range 1-2~M$_\odot$, the $\dot{M}$ of PMS stars in the LMC increases with the stellar mass as 
$\dot{M}  \propto$M$^b _{\star}$ where b$\approx$1, in agreement with the value ($b=0.9\pm0.1$) found for the PMS population of NGC~346 in the SMC (Paper~II).
Taking into account the uncertainty on the $b$-parameter, this dependency is slightly flatter than the second power law 
reported in the same mass range for samples of PMS stars in the Galaxy. 
Using data from the IPHAS survey of PMS stars in the galactic HII region IC~1396, \citet{Bar11} also 
found a slope between 1.2 and 1.3 for the $\dot{M}$ vs. M$_{\star}$ dependency,  
i.e. less steep than claimed in previous literature.

\section{$\dot{M}$ evolution and metallicity dependency}\label{Macc_metal}

The data collected so far for galactic PMS stars indicate that $\dot{M}$ is a decreasing function of 
stellar age \citep{Muz00,Sic05,Sic06,Sic10,Fed10}, roughly in line with the expected evolution of viscous discs \citep{Har98}, although the  spread of the data is very large (Sect.~\ref{intro}). 
More precisely, \citet{Fed10} estimated a mass accretion timescale ($\tau_{\rm acc}$) of  2.3 Myr for galactic PMS stars in the spectral type range K0-M5. 
Moreover, in the pioneering study of the mass accretion process in the MCc presented in Paper~I and Paper~II, we found that, in the field around supernova SN1987A in the LMC 
and in NGC~346 in the Small Magellanic Cloud (SMC), the $\dot{M}$ of PMS stars are systematically higher than their galactic counterparts. 

Because of the very large uncertainties affecting age and $\dot{M}$ determinations, one must be very careful when doing such a comparison for individual young stars. 
In particular, isochronal age estimates for PMS stars in the MCs are more uncertain than ages of galactic PMS stars, 
that can be confirmed by other methods (turnoff of the main sequence, dynamical evolution of the gas, etc.), and star forming regions 
in the MW are usually smaller and, hence, present smaller age dispersion \citep{Elm00}. However, age differences based on statistical samples are more reliable.

In this section, we investigate the $\dot{M}$ evolution in the MCs compared to the MW and verify the result by \citep{DeM10,DeM11}. 
We dispose of an enlarged sample of PMS stars in the LMC with $\dot{M}$ measured in a homogeneous way and, hence, we adopt a robust statistical approach which is expected to 
reduce as much as possible the effect of uncertainties on age/$\dot{M}$. However, there are unavoidable caveats that might affect our results and we will discuss them 
in detail in Sect.~\ref{caveats}.

Figure~\ref{fig_Macc_mass_age} (lower panel) gives a first hint on the time evolution of $\dot{M}$ in the LMC, which, as expected, decreases with age. 
In order to investigate quantitatively the $\dot{M}$ in the LMC we used at first the linear regression fit already describe in Sect.~\ref{Macc_mass_age} (Equation~\ref{regr})
and shown by the solid line in Figure~\ref{fig_Macc_mass_age} (lower panel). We obtain $a \approx$1/3, both when considering all the mass range 1-2~M$_{\odot}$ 
and when limiting the fit to the M$_\star \gtrsim 1.3 M_\odot$ regime to avoid incompleteness effects. 
This would imply that the $\dot{M}$ of PMS stars in the LMC decreases with age significantly more slowly than observed 
for Galactic PMS stars of the same mass \citep[$a \approx$1.2;][]{Sic10}, where the $\dot{M}$ evolution is marginally 
consistent, though still slower, than the prediction of the viscous disk evolution model \citep[a=1.4-2.8;][]{Har98}. 
This comparison is based on age/$\dot{M}$ estimates for single stars and there are at least two main problems affecting its outcome:

\begin{enumerate}

\item In Sect.~\ref{mass_age} we have already pointed out that isochronal ages of individual objects are uncertain, especially below a few Myr, 
and only ages and age differences of statistical samples should be trusted. 

\item \citet{Sic10} already pointed out that the difference between the observed $\dot{M}$ evolution and prediction of the viscous disk 
model might be attributable to the fact that solar-type objects present a variety in the viscosity parameter radial exponent \citep{Ise09} 
and long-surviving disks may be biased toward certain radial viscosity laws. 

\end{enumerate}

\begin{table*}
\caption{Median  $\dot{M}$ and number statistics per age bin for galactic PMS populations in the mass range 1-2~M$_\odot$.}
\begin{center}
\begin{tabular}{lcccl}
\hline\hline
Galactic region  			& Typical Age      & Median $\log{\dot{M}}$ & Number of members with       & Ref.  \\
         					         & (Myr)                  &   (M$_{\odot}$/yr)           &   1$\leq M/M_\odot \leq$2      &           \\
\hline\hline
$\rho$-Ophiuchus$^\dag$   ($\rho$-Oph)    & 0.5    &   -7.95  &   4  & \citet{Nat06}   					\\
\hline
Orion  Nebula Cluster (ONC-young)    			& 3       &   -7.60   &   9  & Panagia et al. 2011 (in preparation)     \\
Orion  Nebula Cluster (ONC-old)$^\ddag$    	         &  12       &   -9.00  &  9   & 							            \\
\hline
IC~348  							& 2-3      &  -7.81  &   9  & \citet{Dah08}  				\\
Lupus~III (LupIII)      					&             &            &        &  \citet{Her08}                                    \\
$\sigma$-Orionis ($\sigma$-Ori)		&             &            &        &  \citet{Rig11}  				\\
Taurus (Tau)						&             &            &        &  \citet{Cal04}  				\\
\hline
Trumpler~37 (Tr37)          				& 4        &  -9.49   &  88 & \citet{Sic06,Sic10,Bar11}                  \\
Chamaeleon~II   (ChaII)  				&            &             &        & \citet{Spe08}     		          \\
\hline
$\eta$-Chamaeleontis ($\eta$-Cha)  	& 5-7       &  -8.98   &  7   & \citet{Jay06}	           \\
Orion OB 1c Association (Ori-OB1c)		&            &                &          & \citet{Cal04}     		          \\								
$\lambda$-Orionis  ($\lambda$-Ori)		&            &               &          & \citet{Cal04}     		          \\	
\hline
Hydrae (Hya)   					         &  8-10      & -10.52 &  4 & \citet{Pin07,Jay06}      \\
\hline
$\beta$-Pictoris ($\beta$-Pic)  			& 12     &     -10.26          &  5   & \citet{Pin07,Sic10}	\\
NGC~7160                                 			&            &                         &        &  \citet{Sic10}                 \\
\hline
Lower Centaurus Crux (LCC) 			&      16     &   -8.32   &  40      &  \citet{Pin07}                 \\
Upper Centaurus Lupus (UCL) 		&              &                &            &  \citet{Pin07}                 \\
\hline
Tucana-Horologium (Tuc-Hor)  		&  30      &     -11.03    &  8        &   \citet{Pin07,Jay06} \\
\hline\hline
\end{tabular}
\label{tab_Macc_Gal}
\normalsize
\end{center}
$^\dag$\footnotesize{Note that, while the age of $\rho$-Oph is often listed as $\sim$0.5~Myr, \citet{Wil08} state that the the well characterized, optically visible
PMS stars might be as old as 2-5~Myr. Thus, in Figure~\ref{Macc_MC_Gal} we indicate with an arrow the possible age range spanned by $\rho$-Oph members.}\\
$^\ddag$\footnotesize{For the ONC-old population we report the minimum age estimated by Panagia et al. (2011, in preparation).}\\
\end{table*}

\begin{figure*}
\includegraphics[angle=0,scale=.60]{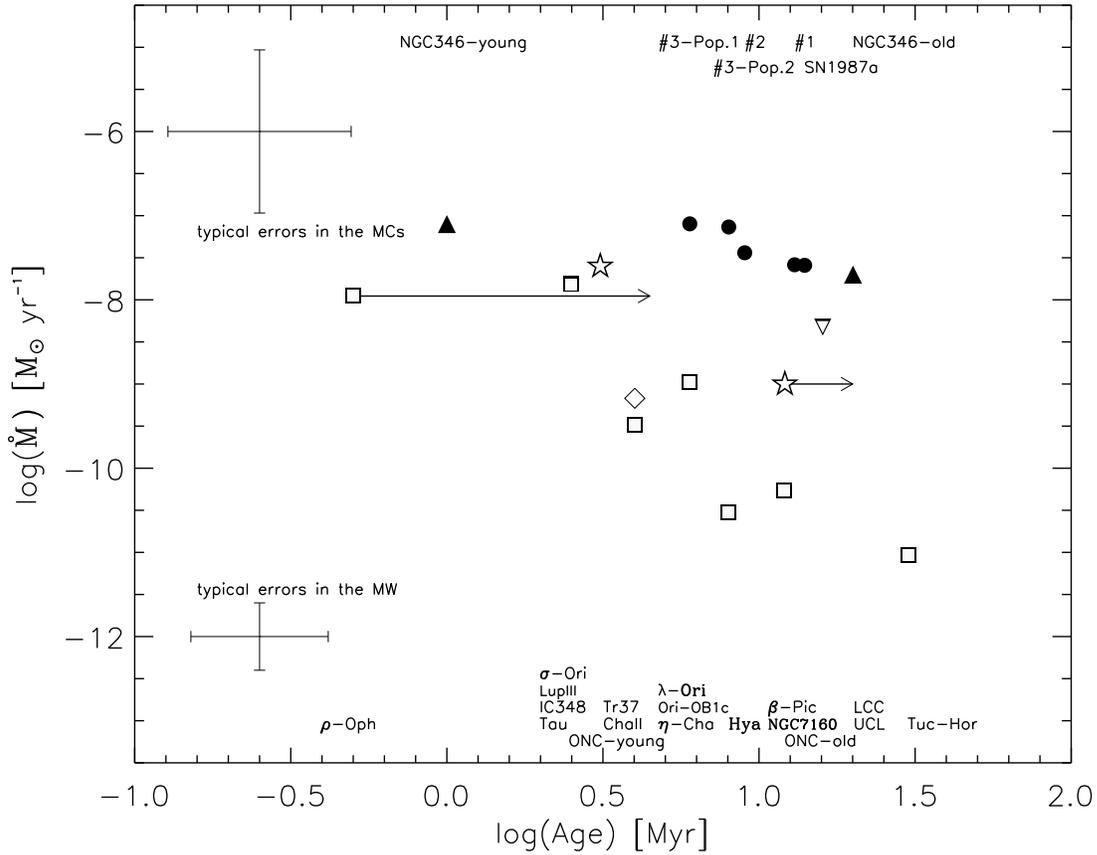}
\caption{Median mass accretion rate as a function of the median age for PMS populations in 
the LMC (filled circles), NGC~346 in the SMC (filled triangles) and 17 young clusters 
and star forming regions in the MW (open squares), as indicated in the labels. The upside-down triangle represents the measurement for the LCC/ UCL region, 
while the diamond and the pentagrams are the photometric measurements by \citet{Bar11} and Panagia et al. (2011, in preparation), respectively. 
The arrows indicate the age uncertainty for the $\rho$-Oph members and the ONC-old population (see Table~\ref{tab_Macc_Gal}).
The typical errors on $\dot{M}$ and age are displayed in the top-left and bottom-left for the MC and MW regions, respectively.}
\label{Macc_MC_Gal}
\end{figure*}

A more robust approach to study the evolution of $\dot{M}$ in the MCs in comparison to our MW consists of deriving the median age and $\dot{M}$ 
of a sample of star forming regions in both galaxies spanning a suitable age range. 
We have collected from the literature 17 Galactic star-forming regions and clusters 
with ages between 1 and 30 Myr whose stellar population is well characterized in terms of $\dot{M}$. 
We considered only Galactic PMS stars with mass in the range 1-2~M$_\odot$, i.e. the range we are studying in the LMC. 
This sample consists of $\sim$170 objects; their typical age and median $\dot{M}$, together with the number statistics per age bin and the reference papers, 
are reported in Table~\ref{tab_Macc_Gal}. 

In Figure~\ref{Macc_MC_Gal} we plot the median $\dot{M}$ as a function of the median age for each one of these 18 Galactic PMS populations, 
for the LMC populations studied in this paper and for  NGC~346 in the SMC \citep[][hereafter Paper~II]{DeM11}. 
Note that NGC~346 displays two different stellar populations with median ages of $\sim$1 (NGC~346-young) and $\sim$20~Myr  (NGC~346-old), 
respectively, and our field \#3 in the LMC also presents two spatially separated PMS populations (\#3-Pop.~1 and \#3-Pop.~2) with slightly 
different median ages (6 and 7~Myr, see Sect.~\ref{OB_photoevap}). Thus, we dispose of 7 regions in the MCs with 
ages between 1 and 20~Myr (Table~\ref{mean_par}). The errors bars in Figure~\ref{Macc_MC_Gal} for the MCs points 
are computed as the typical standard deviation of the $\dot{M}$ and age distribution of each population plus the intrinsic uncertainty on these parameters. 
For the Galactic regions we adopt the typical error on $\dot{M}$ and age estimated by \citet{Sic06}. 

The analysis of Figure~\ref{Macc_MC_Gal} reveals  that below 3~Myr the only information we have at the moment in the 
MCs comes from NGC~346 in the SMC. Moreover, in this age range the number statistics of Galactic PMS 
stars with mass between 1 and 2~M$_\odot$ and measured $\dot{M}$ is very poor (i.e. only $\sim$10\% of the Galactic sample). 
As discussed in Sect.~\ref{caveats}, below 3~Myr age determination becomes critical. Thus, we focus on the age range $>$3~Myrs and 
defer the analysis of the very young regime (1-2~Myr) to a future paper based on improved number 
statistic of 1-3~Myr old PMS stars in the core of the 30~Doradus starburst cluster \citep{DeM11b}. 

After the first 3~Myr, despite the large uncertainties, the average $\dot{M}$ of PMS stars in the MCs is systematically higher 
than that of Galactic PMS stars of the the same age, confirming the results by Paper~I. 
Note that the pentagrams in Figure~\ref{Macc_MC_Gal} display the median $\dot{M}$ for the two PMS populations in the 
Orion Nebular Cluster (ONC-young and ONC-old, see Table~\ref{tab_Macc_Gal}) and the diamond corresponds to Trumpler~37 (Tr~37), 
for which mass accretion rates are photometrically derived by 
Panagia et al. (2011, in preparation) and \citet{Bar11}, respectively, using the same method adopted in this paper. 
These three points follow very well the Galactic trend. In particular, the median $\dot{M}$ measured by \citet{Bar11} for the PMS population 
in Tr~37 perfectly agrees with the median $\dot{M}$ measured by \citet{Sic06,Sic10} for the same population on the basis of U-band excess emission and 
high resolution spectroscopy, which provides accurate spectral type and reddening estimate. 
This demonstrates that the difference in the typical $\dot{M}$ between the MCs and the MW is not due to systematic uncertainties 
of the photometric method we use to measure $\dot{M}$.

If we perform a linear fit to these averaged data for ages  above 3~Myrs, we find that in the MCs $\dot{M}$ decreases 
with time as $\dot{M} \propto t^{-a}$ with $a=1.5\pm0.2$, confirming the average trend 
in the MW ($a =2.6\pm1.0$) and in agreement with the prediction of viscous disk evolution models \citep[a=1.4-2.8;][]{Har98}. 

Being the metallicity of the MCs considerably lower than the MW, Figure~\ref{Macc_MC_Gal} suggests a tantalizing metallicity dependency 
of the mass accretion process, specifically that $\dot{M}$ appears to be inversely proportional to stellar metallicity. 
Under this hypothesis, we should observe some difference in $\dot{M}$ even between the two MCs, 
because their metal content is different \citep[Z=0.007 for LMC and Z=0.002 for SMC;][]{Mae99} and 
the difference is comparable with the difference between the LMC and the MW (Z$\approx$0.018). 
We do not observe such a difference when considering the mean $\dot{M}$ values, as we did in our Figure~\ref{Macc_MC_Gal}. 
However, Paper~II (focused on the the SMC cluster NGC~346 and based on individual age and $\dot{M}$ measurements) 
indicates that there is a difference between the two MCs and $\dot{M}$ in the SMC seems to be slightly higher than in the LMC. 
This would fit the idea of $\dot{M}$ being inversely proportional to the metallicity.

These results suggest that the mass accretion process in the MCs might be different from the typical mechanism observed 
in Galactic PMS stars in two main aspects, which are discussed in the following sub-sections. 
In Sect.~\ref{caveats} we will further discuss these results considering the many caveats of current PMS evolutionary models,  
which might affect age estimates both in the MCs and in our MW and hence, our results. 
We will also review the main mechanisms of disk dispersal (Sect.~\ref{theo}) 
in order to understand whether the lower metallicity of the MCs with respect to the MW plays in favor of a typically higher $\dot{M}$.

\subsection{Disk lifetime}

For PMS stars in the MW most signs of accretion disappear after $\sim$10~Myr. 
This, together with the observation that the fraction of stars with near-IR excess, 
which originate from the hot dust in the inner disk, decreases at a constant rate \citep{Hai01,Sic06}, yields an overall disk lifetime of about 5-6~Myrs. 
However, a few cases of strong accretors at older ages have been reported in the literature. 
\citet{Sic05} notice the presence of several Galactic G-type stars, with an apparent age older than $\sim$10~Myr, showing 
higher accretion rates outside the typical trend of Galactic PMS stars (see their Figure~12). However, they warn that their age estimates might be 
significantly affected by the uncertainties in the birth line for G-type stars (see Sect.~\ref{caveats}). 
More recently, \citet{Bau11} also found a handful of stars with H$\alpha$ emission typical of accretors in the 20-30~Myr old cluster NGC~6167.

Moreover, among the 17 Galactic regions considered in Figure~\ref{Macc_MC_Gal} there is a clear outlyer, represented as an upside-down triangle: 
the 16~Myr old Lower Centaurus Crux (LCC)/ Upper Centaurus Lupus (UCL) region, where the typical $\dot{M}$ is higher than expected 
for its age and is fairly consistent with (though still lower than) that measured for the 20~Myr old population of NGC~346 in the SMC. 
\citet{Pin07} presented a comparative study of the average mass accretion rates in Galactic young clusters and 
associations with ages between 10 and 30 Myr based on photometric excesses in the U band; they already noticed that an important fraction ($\sim$25\%) of stars 
in LCC/UCL presents a relative strong accretion. According to these authors, the peculiar case of LCC/UCL might indicate 
that gas reservoirs in disks can exceptionally exist much longer than 10 Myrs, allowing the formation of giant 
planets up to the age of $\sim$16~Myr \citep[see also][]{Moo11}.  
However, no strong correlation was found between U-band excess emission and H$\alpha$-excess  emission for the stars in the LCC/UCL 
and some doubts arouse about the nature of the U-band excess (R. de la Reza, private communication).

For the sake of clarity, we also point out that the study by \citet{Pin07} finds 8 accretors in Tuc-Hor and 5 in $\beta$-Pic
in the mass range 1-2~M$_\odot$, while previous studies of the PMS population of these associations  \citep{Jay06} 
find only 2 accretors in $\beta$-Pic and  none in Tuc-Hor. The authors do not discuss this discrepancy.

In the MCs, we measure an average $\dot{M}$ of 5$\cdot$10$^{-8} M_{\odot}$/yr for 10-12 Myr old stars, 
indicating that disks in the MCs might be long-lived with respect to the MW.
Observational measurements of the disk lifetime in the LMC and, in general, low-metallicity environments 
are at the moment very poor and controversial. 
Using data from the Spitzer SAGE Survey in the MCs, 
it was found that the fraction of stars with disks in the SMC is higher than in the LMC (D. Lennon, private communication).
Being the SMC more metal poor than the LMC, this would be consistent with a scenario
where the disk lasts longer in a metal poorer environment. 
On the other hand, in the extreme outer Galaxy (EOG), 
where the metallicity is one-tenth of the solar neighborhood, 
\citet{Yas09} found a fraction of stars with circumstellar disks significantly 
lower than in the solar neighborhood, suggesting that most stars in low-metallicity 
environments experience disk dissipation at earlier stages ($<$1~Myr). 
The mismatch between these two results is rather awkward. 
\citet{Yas09} might have underestimate the disk fraction in the EOG because their 
measurements are based only on near-IR excess, which can be very small or inexistent in flattened disks. 
Moreover, \citet{Yas09} warn that a mechanism specific to the 
EOG environment, i.e. effective far-ultraviolet photoevaporation, might contribute to the rapid disk dispersal \citep[see, e.g.,][]{Hay08}.  
The mechanisms of disk dispersal and their dependency on metallicity are further discussed in Sect.~ \ref{theo} to \ref{ext_OB_photoevap}.

\subsection{Disk mass}

In order to keep an average $\dot{M}$ of 5$\cdot$10$^{-8}$~M$_\odot$/yr for $\sim$10~Myr, 
the disks we are observing in the MCs should be very massive, i.e. about 0.5~M$_\odot$. 
Moreover, it is normally assumed that accretion rates are much higher
in the protostellar phase ($\lesssim$1~Myr) than at 5-10~Myr. Thus, if it is 
really true that PMS stars in the LMC have average $\dot{M}$ of 5$\cdot$10$^{-8}$ M$_\odot$/yr  at 10 Myr, 
one would expect them to have even higher $\dot{M}$ at earlier ages and, hence, 
even higher disk mass. 

Disk properties and initial conditions may be different in low-metallicity environments and disk dispersal mechanisms other than accretion 
might contribute to dissipating the disk mass. Moeover, a recent study by \citet{Zhu10} suggests that, if the initial cloud core is moderately rotating, the protostar ends up 
with a more massive disk and most of the central star's mass is built up during FU Ori-like outburst events of accretion. 
However, it is still hard to explain how such massive disks are produced, because oservations in the MW suggest 
that stars accrete most of their mass in the embedded phase, 
and only a few percentage once they reach the T~Tauri phase. 

The high $\dot{M}$ measured in the MCs would imply that the overall accretion process 
during the PMS  phase will add to the central object a mass comparable to that of the central object itself 
and, hence, the protostars itself should have as much mass in the central object as in the disk/envelope. 
This appears unlikely, because the Initial Mass Function (IMF) in the MCs appears remarkably similar to the that of the MW \citep[e.g.,][]{Gar98,Bas10}. 

However, there is an important point to be considered when doing such speculations. 
The mass accretion process of stars in the PMS phase is a highly variable phenomenon 
and this variability is reflected in the diagnostics used to measure  $\dot{M}$ 
\citep[H$\alpha$ excess emission, veiling, mid-IR Br$\gamma$ and Pa$\beta$ emission, etc.;][]{Her86,Har91,Ngu09,Fed10,Pet11,Fae11}. 
At any given time, only a certain fraction of PMS stars of the given PMS population shows signatures of mass accretion above the detection limit of the given survey.
The study by \citet{Fed10}, based on H$\alpha$ excess emission measurements, shows that the fraction of PMS 
stars with $\dot{M}$ above their detection limit in Galactic star forming regions  decreases with age. 
At the typical age of our PMS star candidates ($\sim$10~Myr), the fraction of PMS stars with $\dot{M}$ above their detection limit is expected to be of the order of  5\%
 \citep[see Figure~3 and 4 by][]{Fed10} or, in other words, their $\dot{M}$ is above the detection limit only for 1/20 of the time. 

Since the mass accretion process depends on metallicity (see Sect.~\ref{theo}), this estimate might  not be accurate for low-metallicity PMS stars. 
Indeed, \citet{DeM11c} found that the fraction of PMS stars above their $\dot{M}$ detection limit at ages younger than $\sim$10~Myr 
in the SMC cluster NGC~346 is of the order of 30\%. In other words, 
these PMS stars are  above the $\dot{M}$ detection limit only for 1/3 of the time.
The metallicity of the LMC is in between the SMC and the MW; we estimate that our PMS star candidates in the LMC, with an average $\dot{M}$ of 5$\cdot 10^{-8}$~M$_\odot$/yr, 
a typical age of $\sim$10~Myr, accrete a total mass of $\sim$0.02 up to $\sim$0.2~M$_\odot$ depending on whether we assume that their are 
above our $\dot{M}$ detection limit (dashed line in Figure~\ref{fig_Macc_mass_age}) 1/20 or  1/3 of the time.
This implies that the disk mass is between 10 and 20\% of the stellar mass, 
because our PMS star candidates have mass in the range 1-2~M$_\odot$,  
which is plausible \citep{Whi04,Alc08,Kim09}.

\section{Caveats on our age and $\dot{M}$ estimates}\label{caveats}

As mentioned in Section~\ref{mass_age},  the uncertainty on stellar masses and ages derived from PMS evolutionary models 
could have a significant impact on our results. In particular, a few caveats on isochronal ages should be taken into account.

Recently \citet{Bar09} showed that, when vigorous episodes of mass accretion at early stages of the stellar life ($\lesssim$1~Myr) 
are taken into account in the calculation of  PMS evolutionary tracks, protostars may have much smaller radii than found in previous treatments. 
Such small radii would have the effect of making some young stars appear fainter and thus much older in the HR diagram, perhaps as much as 10 Myr, than they really are. 
This means that the concept of stellar birth-line has no valid support for very low-mass stars because it depends 
on the accretion history. As a consequence, ages below a few Myr are highly uncertain and 
absolute ages are hard to determine, though relative ages are still reliable.  
This problem has a strong impact on $\dot{M}$ evolution studies, because it implies that a strongly accreting object might show a position 
on the HR diagram consistent with a 10~Myr age, even though its true age is much younger. 

However, the prediction by \citet{Bar09} applies to the limit of low-temperature ("cold") accretion, and 
\citet{Har11} argue that very rapid disk accretion is unlikely to be cold, for several reasons:
i) the FU Ori objects, e.g. the best-studied PMS disks with rapid accretion outbursts, 
have spectral energy distributions consistent with large, not small, radii; 
ii) theoretical models indicate that at high $\dot{M}$, protostellar disks become internally hot and geometrically thick, 
making it much more likely that hot material is added to the star; 
iii) the luminosity of the accretion disk is likely to irradiate the central star strongly, heating up the outer layers and potentially expanding them. 
iv) observed HR diagram positions of most young stars are inconsistent with the rapid cold accretion models. 

The \citet{Bar09} prediction implies that we might have overestimated the age of our PMS star candidates. 
This problem affects age estimates both in the MCs and in our MW; however, under the hypothesis that 
the typical $\dot{M}$ are higher in the MCs, it is expected to be more significant for PMS stars in the MCs. 
In other words, if we accept a typically higher $\dot{M}$ in the MCs, then we have to accept that our age estimates might be inconsistently derived, 
because isochronal ages are not robust for strongly accreting objects and we might have overestimated the age of PMS stars in the MCs. 
This problem can be properly addressed only when improved isochrones for low-metallicity PMS stars will be available. 
However, this issue does not affect considerably our conclusion based on Figure~\ref{Macc_MC_Gal}, 
because the \citet{Bar09} effect is expected to be more significant for very young objects ($\lesssim$1~My) 
of spectral type M or later, ranging from a few Jupiter masses to a few tenths of a solar mass,  
while our sample consists essentially of G and K type stars with ages older than $\sim$5~My (see Table~\ref{mean_par}). 

\citet{Har03} discussed the uncertainties in the birth line of G-type stars (T$_{eff} \gtrsim$5400~K), which results 
in a overestimate of their ages and dominate the apparent age spread. 
We do observe that G-type star in our sample are systematically older than K-type stars by a few Myr. This age difference is partly explained by the fact that our photometric 
limits (Table~\ref{obs_tab}) allow us to detect only the brightest and, hence, youngest late K-type stars at the distance of the LMC, 
while they are sufficient to detect G-type stars of any age and spectral sub-class up to the main sequence (see also Figure~\ref{fig_Macc_mass_age}, upper panel). 
However, we can not exclude some residual effect due to the uncertainty in the birth line of G-type star. This would point toward an overestimation 
of the typical age of PMS populations in the MCs.

Finally, in Sect.~\ref{mass_age} we pointed out that the regions we are studying are very large, about 100~pc in diameter, and we expect an age dispersions of a 
few to 10 Myr within each region (Table~\ref{mean_par}). This age dispersion results from the presence of multiple stellar populations in the same region, 
which might be the result of sequential and/or triggered star formation \citep{Elm00}. A typical example in our MW is the Cep~OB2 bubble, 
which is $\sim$50~pc in diameter and includes the 12 Myr old cluster NGC~7160, the 4 Myr old Tr~37, the~1 Myr old Tr~37 globule, 
the 2-3 Myr old Cep~B, and several other star formation episodes with ages ranging 1-3 Myr \citep{Pat98,Sic05}. 

Similarly, the three LMC regions analyzed in this work present a significant age spread.  Figure~\ref{HR_diag} shows the HR diagram for the PMS population of these regions 
and indicates that they have a typical age of $\sim$13~Myr, $\sim$9~Myr and 6-8~Myr for region \#1, \#2 and \#3, respectively; 
however, a small fraction of the PMS population in each region is as young as 1~Myr. 
Since it is normally assumed that accretion rates are much higher at ages around 1~Myr than at 5-10~Myr \cite[see, e.g., Figure~2 by][]{Sic10}, 
the median $\dot{M}$ we measure could be driven by the younger populations present in each region.
This problem affect also the age and $\dot{M}$ estimates for the MW regions, but it is expected to be less significant because these regions are much smaller. 

On the other hand, our conclusion in Sect.~\ref{Macc_metal} are driven by Figure~\ref{Macc_MC_Gal} using the median age and $\dot{M}$ of each region in the MCs and the MW.
This statistical approach is expected to smoothen the age uncertainty due to strong accretors and/or G-type stars and, indeed, 
the median isochronal ages estimated for the LMC regions are supported by several arguments and independent age estimates (Sect.~\ref{mass_age}). 
However, considering all the caveats listed above, it could be that we are slightly overestimating the age of  the MCs regions. 
Under the most conservative hypothesis that the MC regions are a few Myr younger than our estimates, 
we still find a systematic and interesting differences between the MCs and the MWs, as the accretion
rates in the MCs are about an order of magnitude higher than in Galactic regions of similar age. 
As we will see in Sect.~\ref{theo}, this difference can be justified on the basis of the different metallicity between the MCs and the MW.

\section{Mechanisms of disk dispersal: why low-metallicity disks may have higher accretion rates?} \label{theo}

Our pioneer study of the mass accretion process in the LMC suggests that disks around metal-poor stars accrete 
at very high rates with respect to galactic PMS stars with roughly solar metallicity. 
The many caveats of age/$\dot{M}$ measurements do not allow us to quantify such a difference with high accuracy. 
However, all our arguments support a difference of at least an order of magnitude in $\dot{M}$. 
Models of disk evolution provide several explanations both for and against this observational evidence. 
In the next sections we review the interplay among the main mechanisms of this dispersal 
in low-metallicity environments and in relation to our finding in the LMC.

\subsection{Viscous accretion}  \label{visc}

Viscous accretion is the mechanism by which mass is accreted onto the central star 
because the transport of angular momentum outward allows disk material to flow radially inward \citep{Hol00}. 
In general, a lower metallicity implies a lower disk opacity, lower disk temperature, lower viscosity, 
and thus longer viscous time \citep[see, e.g., review by][]{Dur07}. 
Therefore, it is expected that low-metallicity stars will undergo significant accretion up to older ages with respect to higher metallicity stars, 
in general agreement with our measurements in the LMC. 
However, a few warnings must be taken into account:

\begin{enumerate}

\item According to \citep{Har06}, in disk around low-mass stars the accretion/viscosity is mainly driven by magneto-rotational instability (MRI), which may be more
efficient in low metallicity stars. Indeed, this is the argument used by \citet{Yas09} to explain the fact that in the EOG, where the metallicity is one-tenth of the solar neighborhood, 
the fraction of stars with circumstellar disk is significantly lower, implying a faster disk dispersal. 
This consideration would go against he long-lived low-metallicity disks hypothesis. 
Nevertheless, there is no observational evidence that accretion/viscosity driven by MRI is so efficient.  
For example, evolved disks with large grains, which are expected to behave like low-metallicity disks, accrete less as they age, 
while, under the assumption that the MRI is more efficient at low-metallicity, we would expect the opposite \citep[see Sect.~5 by][and references therein]{Har06}.

\item In order to keep an average accretion rate of 10$^{-7}$ M$_\odot$/yr going on for 5-10 Myr, 
the disks we are observing in the LMC should be extremely massive, about 0.5-1 M$_\odot$. 
An additional effect to take into account in massive disks is gravitational instability (GI). 
It has been suggested that lower opacity favor GI \citep{Cos09} and that low metallicity disks may be more unstable
against GI \citep{Cai06}. There is an ongoing debate about whether GI makes very massive disks fragment \citep{Bos97,Bos03,Cai06,Cai09},
or simply favor accretion and mass transport \citep{Har06}. In both cases, the GI global effect would point against the long-lived low-metallicity disk hypothesis, 
even though it could explain the large $\dot{M}$ we measure.

\end{enumerate}

\subsection{Grain growth} 

According to the analysis by \citet{Dul05}, the growth of dust particles in the disk depends on the disk's dust to gas ratio. 
Low-metallicity disks have a lower dust content and, hence, the process is slower and the grains remain small.
Grain growth is an important contributing factor to disk dissipation by planet formation and photo-evaporation, which we will analyze later on.
In addition, if the disk is less optically thick, the convective turbulence is reduced, making the grain coagulation/fragmentation process less efficient. 
In other words, low-metallicity disks may reach a quiet steady state characterized by small grains that cannot coagulate and,  hence,  they 
last longer, because dissipation by planet formation and/or photo-evaporation is less efficient.  
This scenario is in agreement with our measurements in the LMC, but again there are a few warnings to take into account:

\begin{enumerate}

\item Convection is not the only source of turbulence. As seen in Sect.~\ref{visc}, MRI driven turbulence will be present in the magnetically active parts of the disk. 
However, the high $\dot{M}$ we measure at older ages suggests large disk mass to sustain such long accretion lifetimes and, hence, 
a high gas column density. This kind of disks would present magnetically ``dead zones'' \citep{Tur07,Tur08} with lower turbulence.
These dead zones might suffer from other types of instability \citep[e.g. streaming instability,][]{Joh07} and they may be good locations for planetesimal formation. 
However, as seen in Sect.~ \ref{Macc_metal} and \ref{visc}, such large disk masses are difficult to explain.

\item The lower dust content of low-metallicity disk implies that the gas reservoir of the disk is less shielded against photo-evaporation and it might depleted faster. 
Disk photo-evaporation effects and their dependency on metallicity are discussed in Sect.~\ref{OB_photoevap}.

\end{enumerate}

\subsection{Planet formation} 

\citet{Joh09} presented three dimensional numerical simulations of particle clumping and planetesimal formation in protoplanetary disks with varying amounts of solid material.
These simulations showed that the formation of planetesimals is strongly dependent on the dust to gas ratio. 
Low-metallicity disks would not be able to produce planets at the beginning, but they may do it later on once other mechanisms, 
like internal photo-evaporation, have reduced the amount of gas. 
Thus, the correlation between host star metallicity and exoplanets may reflect the early stages of planet formation and 
low-metallicity disks can be particle enriched during the gas dispersal phase, leading to a late burst of planetesimal formation. 
The metallicities studied by \citet{Joh09} are solar or super-solar, not as low as in the LMC. 
However, if low-metallicity disks would produce planets only in late phases, they might be able to keep an average $\dot{M}$ higher than 
higher metallicity disk, because early proto-planet growth might reduce the accretion to the central star \citep[][and references therein]{Mer10}. 
In a very low-metallicity environment such as the LMC, this effect might be enhanced and would explain  the large $\dot{M}$ we measure.

For the sake of completeness, we defer the reader to the following works pointing out other open issues and possible bias on the planet-metallicity correlation:

\begin{itemize}

\item \citet{Yas09} found that most stars in low-metallicity environments experience disk dissipation at earlier stage ($<$1~Myr) than in the solar neighborhood and 
argue that this could be one of the major reasons for the strong planet-metallicity correlation (see Sec.~\ref{intro}), 
because the shorter disk lifetime reduces the time available for planet formation. This result is in disagreement with our measurements for low-metallicity PMS stars in the LMC, 
whose accretion process appears to last longer than for galactic PMS stars.
In Sect.~\ref{Macc_metal}, we have seen that the results by \citet{Yas09} might suffer incompleteness effects and the rapid disk dispersal in the EOG  might be due to 
effective far-ultraviolet photo-evaporation rather than planet formation. 
Moreover, \citet{Erc10a} explored the metallicity dependence of a dispersal mechanism based on planet formation in the core accretion scenario \citep{Ida04} 
and found opposite results, i.e. planet formation in low-metallicity environments would extend disk lifetime, 
because the lower metal content implies a smaller amount of dust and, hence, a slower planet formation process. 
These authors explain the observations in the EOG in the context of a 
disc dispersal mechanism based on X-ray photo-evaporation from the central star, 
which operates faster in low metallicity environments (see discussion in Sect.~\ref{OB_photoevap}).

\item According to \citet{Bou09}, planets formed in lower-metallicity disks would accrete less gas and would not suffer so much migration; 
thus, they may end up as low-mass planets very hard to detect.

\end{itemize}

\subsection{On the disk dispersal through photo-evaporation}\label{OB_photoevap}

Photo-evaporation of the surface layer of the gas in the disk can be triggered by strong radiation fields generated by nearby OB stars (external photo-evaporation) and/or 
the radiation field of the central PMS star itself \citep[internal photo-evaporation][]{Hol00}. 
These two mechanisms operate very differently and have very different responses to changes in metallicity. Thus, we analyze them separately.

\subsubsection{Internal photo-evaporation} \label{int_OB_photoevap}

Internal photo-evaporation is the mechanism by which radiation from the central source heats the gas in the disk, enabling it to climb up the gravitational potential well and escape. 
Theoretical studies of the erosion of circumstellar disks  around solar-type stars via internal photo-evaporation suggest typical mass loss rates of $\sim10^{-10} M_{\odot}/yr$ \citep{Ale06a,Ale06b} 
for photo-evaporation caused by extreme ultra-violet (EUV) radiation and $\sim10^{-8} M_\odot/yr$ for photo-evaporation caused by X-rays \citep{Owe10}. 
Thus, X-rays appear to be much more efficient at driving internal photo-evaporation \citep[see discussion in][]{Erc09,Erc10b}. 
\citet{Erc10a} explored the metallicity (Z)  dependence of X-ray mass loss rates and found that X-ray wind rates are higher in lower-Z environments and, consequently, 
disk lifetimes are shorter. This occurs because heavy elements represent the main source of opacity for X-rays; the low metallicity implies a reduction of the X-ray opacity and, 
hence, a larger column can be heated by X-rays \citep[see Figure~2 of][]{Erc10a} driving a denser wind from deeper down in the disk\footnote{Note that photo-evaporation driven by EUV or far ultra-violet (FUV) radiation will not give this result. In the case of FUV, the lack of metals  and, hence, dust would actually hinder 
heating although the reduction of opacity is still present \citep[][]{Gor09}, while in the case of EUV radiation the main source of opacity is Hydrogen rather than metals.}. 
We find that metal-poor stars in the LMC accrete at very high rates with respect to galactic PMS stars of the same age, which can be explained in the context of the models presented by \citet{Erc10a}. 
These authors provides preliminary evidence for shorter disk lifetimes at lower metallicities when X-ray driven photo-evaporation is considered as the main disk dispersal mechanism. 
To a first approximation a disc will be destroyed by photo-evaporation when $\dot{M}$=$\dot{M}_W$(Z), where $\dot{M}$ is the 
 accretion rate through the disk and $\dot{M}_W$(Z) is the X-ray wind rate. \citet{Erc10a} show that $\dot{M_W} \propto Z^{-0.77}$. 
 This dependency is steep enough to compensate for the about one order of magnitude difference in $\dot{M}$ 
between the LMC and the MW, because the LMC has about half solar metallicity. 
Moreover, mass accretion rates al lower metallicity are expected to be higher because the disk viscosity parameter ($\alpha_{acc}$) is higher at lower Z. 
In the MRI scenario this occurs because, thanks to the lower X-ray opacity, a larger mass fraction of the disk is ionized and therefore MRI active.

Thus, a disk dispersal scenario based on internal X-ray photo-evaporation is consistent with the idea of higher $\dot{M}$ at lower metallicity.

\subsubsection{External photo-evaporation} \label{ext_OB_photoevap}

External photo-evaporation, caused by nearby massive stars, works primarily via FUV heating of the gas in the outer disk and relies on photoelectric heating from dust grains. 
The most striking example of external photo-evaporation is seen in the proplyds in Orion \citep{Bal98,Bal00}. \citet{Lin07} also argued that 
the ultra-violet (UV) emission and winds of nearby hot stars play an important role in the development of evaporating gaseous globules \citep{Hes96} into protostars. 
External photo-evaporation is difficult to quantify due to a number of uncertainties, including projection effects, and there still is no clear indication of its global effect on PMS populations.

To study the effect of external photo-evaporation on the PMS populations in our fields, we analyze the spatial distribution of the 
PMS stars and their level of H$\alpha$ luminosity with respect to the distance from the hot OB stars in the field, following the same approach as in Paper~I. 
To identify OB stars in our fields, we used the catalog of UBVI stellar photometry of bright stars  across the central 64~deg$^2$ area 
of the LMC and the relative extinction map by \citet{Zar04}. We compare the dereddened photometry of these bright stars 
with the tabulation of absolute magnitudes of OB stars by \citet{Weg00}, assuming a distance modulus of 51.4$\pm$1.2~kpc \citep{Pan91,Pan91b}.
Only hot stars with spectral type earlier than B1 were considered, because cooler 
stars do not contribute significantly to the ionization (see Table~1 in Paper~I).
The center of mass of the hot stars in each field was  computed using as weight their bolometric luminosity.

\begin{figure}
\includegraphics[angle=0,scale=.425]{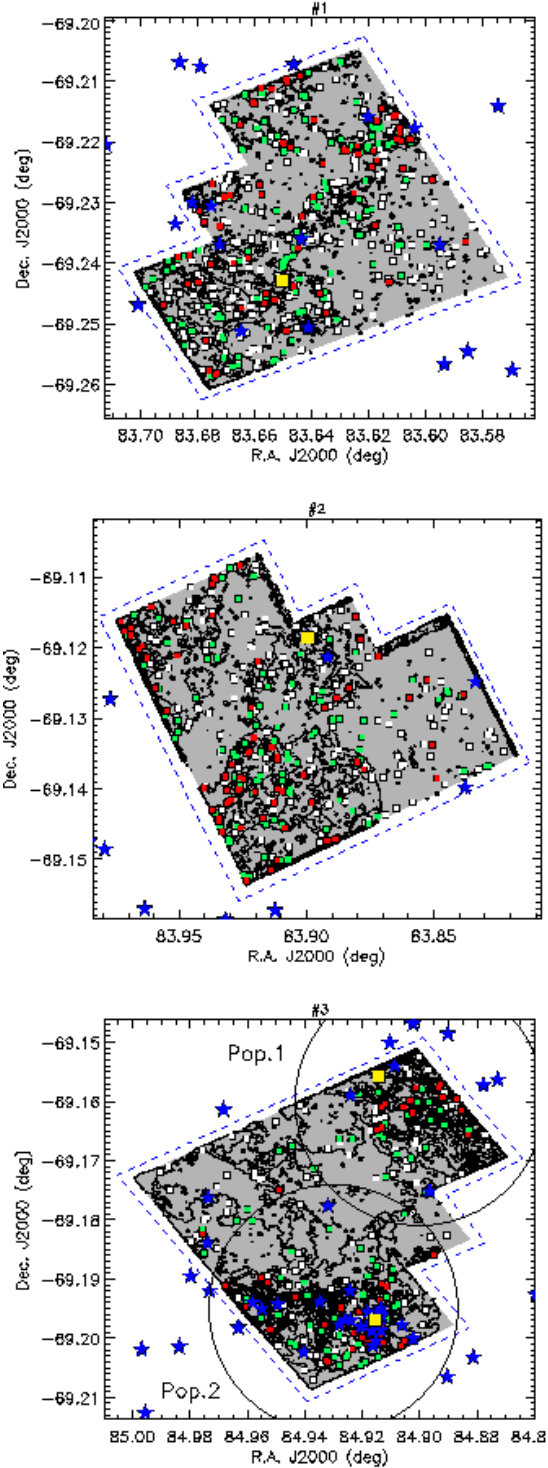}
\caption{Spatial distribution of the PMS stars in our LMC fields: red squares correspond to PMS stars with L(H$\alpha$)$> 0.03 L_\odot$, 
white squares indicate objects with L(H$\alpha$)$< 0.01 L_\odot$, and green diamonds are used for intermediate values. 
The blue stars indicate the position of the ionizing stars in each field and the large yellow square indicates the relative center of mass. 
The contours of the H$\alpha$ nebular emission are also drawn (black lines). 
The two big circles in the lower panel indicate the two possible PMS populations in field \#3 and the two yellow squares 
show the position of the center of mass of the ionizing stars for these two populations.}
\label{spa_dist}
\end{figure}

Figure~\ref{spa_dist} shows the spatial distribution of the PMS stars in our LMC fields and the position of the ionizing stars. 
The center of mass of the ionizing flux is represented in each region by the large square. Symbols of different colors indicate PMS stars with 
different level of H$\alpha$ luminosity, as indicated in the caption, and the lines display the contours of the H$\alpha$ nebular emission 
retrieved from our WFPC3/f656n images. Young stars tend to concentrate in region of high dust/gas density and indeed, 
most of the PMS stars in our sample follow the lanes of the nebular emission. We note, in particular, that our field \#3 presents 
two concentrations of PMS stars corresponding to two well defined over-densities in the nebular emission. 
Moreover, one of the two groups is also associated to a clear cluster of OB stars. We do not observe a significant age difference between 
these two groups (Table~\ref{mean_par}), also because of the limitations in the age determination (Sect.~\ref{mass_age}). 
However, the peculiar spatial distribution suggests that two populations of 
PMS stars separated by about 0.035~deg (i.e. $\sim$30~pc at the distance of the LMC) 
might be present in this region, as indicated by the circles in Figure~\ref{spa_dist}. 
Thus,  in our analysis we deal with these two groups separately and refer to them as field \#3-Pop.1 and field \#3-Pop.2.

\begin{figure}
\includegraphics[angle=0,scale=.35]{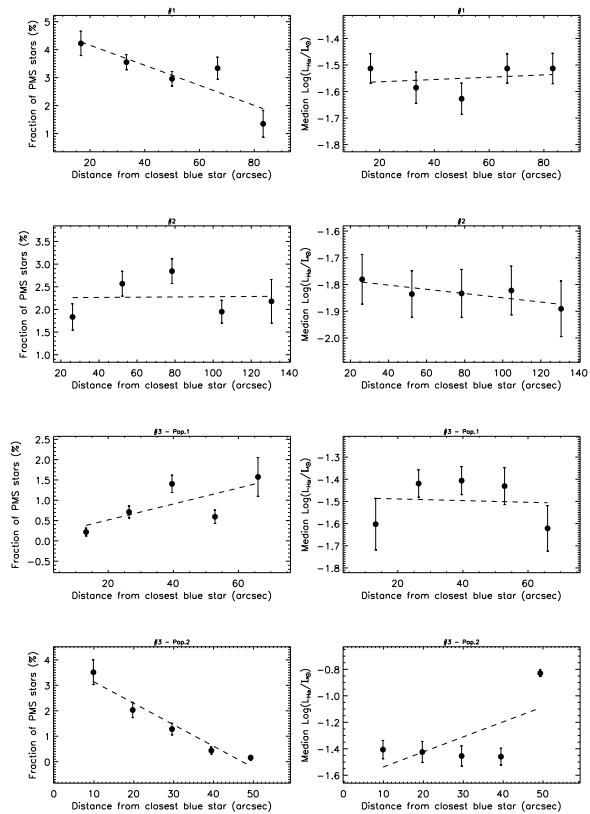}
\caption{Fraction of PMS stars with respect to all stars (left panels) and relative median H$\alpha$ luminosity (right panels) as a function of the distance 
from the closest ionizing star for our fields \#1, \#2 and the two possible PMS populations in field \#3 in the LMC.  
The dashed line in each panel is the linear fit to the points. }
\label{OB_stars_dist}
\end{figure}

Figure~\ref{spa_dist} shows a variety of behaviors in the spatial distribution and typical H$\alpha$ luminosity 
of the PMS stars with respect to the center of mass of the OB stars for the three LMC fields. 
To show this difference quantitatively, we plot in Figure~\ref{OB_stars_dist} the frequency of PMS  stars and their typical H$\alpha$ luminosity 
as a function of the distance to the closest ionizing OB stars for 5 distance bins. The dashed lines display the linear fit to the points, 
which gives and indication of the global trend of the data. The same plots have been presented in Paper~I 
for the field around the supernova SN1987A (see their Figure~13 and 14). 
The result of this analysis for the LMC fields can be summarized as follows.

For the field around supernova SN1987A, we found in Paper~I that PMS stars in the vicinity of OB stars 
are both less numerous and fainter in H$\alpha$ emission than farther away. 
This suggests that their circumstellar disks have been considerably eroded, and made less efficient, by enhanced photo-evaporation 
caused by the ionizing radiation of the massive stars integrated over their $\sim$2~Myrs lifetime. 
As shown in Figure~\ref{OB_stars_dist}, we find a similar behavior in our field \#3-Pop.1, where the number 
of PMS star candidates in the vicinity of OB stars is lower than farther away. 
However, the H$\alpha$ luminosity of this PMS population does not show any specific trend with respect to the position of OB stars.

The inspection of Figure~\ref{OB_stars_dist} for the PMS population of our field \#1 and field \#3-Pop.2 shows a different trend. 
In these two fields there are more PMS stars near the ionizing OB stars, while their typical H$\alpha$ luminosity is roughly constant. 
These correlations are consistent with a scenario where photo-evaporation due to ionizing radiation of nearby OB stars does not play a 
significant role for disk dissipation.

Moreover, the fact that the distribution of low-mass stars follows the distribution of massive stars 
suggests that i) either they belong to the same generation or ii) they formed during separate events of star formations sharing the same 
center of mass. Regions \#1 and \#3-Pop.2 are 11 and 6~Myr old, respectively (Table~\ref{mean_par}); 
although we could have slightly overestimated the age of these regions (Sect.~\ref{caveats}), both of them are too old to assume 
that the low-mass PMS stars and the OB stars belong to the same generation. 
Thus, our data support hypothesis ii), i.e. two events of star formation separated by a few Myr might have occurred in these regions. 
Indication of multiple events of star formation within the same cluster have been already reported in the literature \citep{Pat98,Sic05,Vin09,Mil09,Bec10,DeM11}. 

The inspection of Figure~\ref{spa_dist} and \ref{OB_stars_dist} for field \#2 does not reveal any clear trend. 

Thus, the LMC stellar populations show a variety of behaviors with respect to disk photo-evaporation due to external radiation fields.
Although this result might appear puzzling, there is a number of considerations which help illustrating the complexity of the problem and understanding 
the ``no trend" detection in field \#2. Indeed, even in our Galaxy, it is very common to see disk photo-evaporation effects in individual case 
\citep[see, e.g., the case of Trumpler~37;][]{Mer09}, while finding a ``global" trend is much harder for several reasons.

The first reason in purely statistical. Even in a large, massive cluster there will be few objects close enough to the massive
stars ($\sim$1~pc) to significantly suffer photo-evaporation effects. In addition, we can only measure projected distances and this implies that 
lots of contaminants, that are not as close to the OB stars as we might think, are included in the analysis. 

Secondly, low-mass PMS stars need to spend a few Myr within 1-2~pc  distance from a OB star to significantly 
suffer disk photo-evaporation effects, though the timescale of this process is still very uncertain. With a typical dispersion velocity of 1-2 km/s 
\citep[see, e.g.,][and reference therein]{Kra08},  many low-mass stars move 1-2 pc per Myr or more if they are ejected and, hence, 
even if they were originally close to a massive star they may end up somewhere else.

Furthermore, the structure/orientation of the given young cluster can be such that it happens to have another young population in the close foreground/background. 
This second population is not physically connected with the cluster under study, but it is indistinguishable because we can only measure projected distances. 
In these cases, any correlation with the position of OB stars would be even harder to detect. 

Finally, triggered star formation may mimic the effect of external photo-evaporation. 
If the winds of a massive stars triggers more star formation in the outskirts of the cluster, as seen in the galactic cluster Tr ~37 \citep{Sic05}, 
the disk fraction is higher at larger distances from the massive star. This occur just because the stellar population there is younger and is not a disk photo-evaporation effect.

Because of these overall uncertainties, we consider it premature to discuss the metallicity dependency of disc dispersal through external photo-evaporation.

\section{Conclusions} \label{conclu}

We presented a multi-wavelength study of three star forming regions in the LMC spanning the age range 1-14~Myr.
We aim at investigating the mass accretion process in an environment with a metallicity significantly lower than our MW. 
We identify about 1000 PMS star candidates actively undergoing mass accretion in these regions on the basis of their H$\alpha$ emission and estimate their mass, age and 
mass accretion rate. Our measurements represent the largest $\dot{M}$ dataset for low-metallicity stars presented so far. 
The conclusions of our study are as follows:

\begin{itemize}

\item In the mass range 1-2~M$_\odot$, the $\dot{M}$ of PMS stars in the LMC increases with stellar mass as 
$\dot{M}  \propto$M$^b _{\star}$ with $b \approx$1, i.e. slightly slower than the second power low 
previously reported for galactic PMS stars in the same mass regime, and in agreement with the recent results
by \citet{Bar11} for the galactic HII region IC~1396 ($b$=1.2-1.3).

\item We find that the typical $\dot{M}$ of PMS stars in the LMC is higher than for galactic PMS stars of the same mass, independently of their age. 
Considering the caveats of isochronal age and $\dot{M}$ estimates, the typical difference in $\dot{M}$ between the MCs and our MW 
appear to be about an order of magnitude;

\item Currently available models of disk evolution/dispersal support the hypothesis that the higher $\dot{M}$ 
measured in the LMC might be a consequence of its lower metallicity with respect to our MW;

\item The sample of PMS populations studied in the LMC shows a variety of behaviors with respect to disk photo-evaporation due to external radiation fields. 
In the region around supernova SN~1987A (Paper~I) we found clear evidence that circumstellar disks have been eroded by photo-evaporation caused by nearby massive stars. 
In the three regions analyzed in this work, there is no clear evidence of such an effect. However, the analysis of the spatial distribution 
of the PMS populations in region \#1 and \#3-Pop.~2 with respect to the massive OB stars in the field revealed signs of sequential star formation.

\end{itemize}

In order to clarify these issues, improved evolutionary models for low-metallicity PMS stars and modeling/calculations of disk dispersal 
at low-metallicity are needed, as well as observations of slightly older (e.g. 30-50~Myr) low-metallicity regions than those presented in this paper.

\section*{Acknowledgments}

We are grateful to an anonymous referee whose suggestions have helped us to improve the presentation of this work.
We thank A. Natta, L. Testi and G. Beccari, for the many useful discussions and suggestions, and M. Romaniello for making available his tool for calculating stellar masses and ages. 
This publication makes extensive use of data products  from the HST archive at the Canadian Astronomy Data Centre (CADC), 
operated by the National Research Council of Canada with the support of the Canadian Space Agency.

\bsp

\label{lastpage}

\end{document}